\documentclass[smallextended]{svjour3}
\usepackage{graphicx, amsmath, booktabs, multirow}
\usepackage[caption=false]{subfig} 

\newcommand{\ket}[1]{{\left| #1 \right\rangle}}

\journalname{Quantum Inf Process}

\begin{document}

\title{Spatial Search by Continuous-Time Quantum Walk with Multiple Marked Vertices}
\titlerunning{Spatial Search by Quantum Walk with Multiple Marked Vertices}

\author{Thomas G.~Wong}

\authorrunning{T.~G.~Wong}

\institute{T.~G.~Wong \at
	   Faculty of Computing, University of Latvia, Rai\c{n}a bulv.~19, R\=\i ga, LV-1586, Latvia \\
	   \email{twong@lu.lv}
}

\date{Received: date / Accepted: date}

\maketitle

\begin{abstract}
	In the typical spatial search problems solved by continuous-time quantum walk, changing the location of the marked vertices does not alter the search problem. In this paper, we consider search when this is no longer true. In particular, we analytically solve search on the ``simplex of $K_M$ complete graphs'' with all configurations of two marked vertices, two configurations of $M+1$ marked vertices, and two configurations of $2(M+1)$ marked vertices, showing that the location of the marked vertices can dramatically influence the required jumping rate of the quantum walk, such that using the wrong configuration's value can cause the search to fail. This sensitivity to the jumping rate is an issue unique to continuous-time quantum walks that does not affect discrete-time ones.
\end{abstract}


\section{Introduction}

Schr\"odinger's equation \cite{Schrodinger1926,Sakurai1993} is the fundamental equation of quantum mechanics that describes the evolution of a quantum state $\psi$ in continuous time:
\[ i \frac{\partial \psi}{\partial t} = H \psi. \]
We have set $\hbar = 1$, and $H$ is the Hamiltonian that characterizes the total energy of the system. For a particle of mass $m$ in free space, the Hamiltonian is simply the kinetic energy operator:
\[ H = \frac{-1}{2m} \nabla^2, \]
where $\nabla^2$ is the Laplace operator.

The continuous-time quantum walk \cite{FG1998b} is simply the discrete-space analogue of this. That is, the particle is confined to discrete positions in space, which can be expressed as the vertices of a graph, and to transitions expressed as the edges of the graph. Then the Laplace operator $\nabla^2$ is replaced by its discrete version $L = A - D$, where $A$ is the adjacency matrix of the graph ($A_{ij} = 1$ if two vertices $i$ and $j$ are adjacent, and 0 otherwise), and $D$ is the diagonal degree matrix ($D_{ii} = \text{deg}(i)$). With this discrete substitution, the Hamiltonian for a continuous-time quantum walk is
\[ H_\text{walk} = -\gamma L, \]
where we have also grouped the coefficients together into $\gamma$, which is the jumping rate (\textit{i.e.} amplitude per time) of the walk. Evolution by Schr\"odinger's equation with this Hamiltonian is precisely the definition of a continuous-time quantum walk \cite{FG1998b,CG2004}.

As an algorithmic tool, continuous-time quantum walks have been used for a variety of applications, including evaluating boolean formulas \cite{FGG2008}, identifying graph isomorphism \cite{Rudinger2012}, and performing universal computation \cite{Childs2009}. In this paper, we focus on their application to search \cite{CG2004} on regular graphs. This leads to two changes to the Hamiltonian. First, since we assume that the graph is regular, the degree matrix $D$ is simply a multiple of the identity matrix, so it can be dropped by rezeroing the energy or by factoring out a global, unobservable phase. Thus we can use the adjacency matrix $A$ instead of the graph Laplacian $L$. Second, we introduce a term that acts as a ``Hamiltonian oracle'' \cite{Mochon2007}, which marks the $k$ vertices to search for by potentials. With these two changes, the search Hamiltonian is
\begin{equation}
	\label{eq:H}
	H = -\gamma A - \!\!\!\!\!\! \sum_{i \in \text{marked}} \!\!\!\!\!\! | i \rangle \langle i |.
\end{equation}
The goal is to evolve the system $\ket{\psi(t)}$ by this Hamiltonian for as little time possible, from initially being in an equal superposition $\ket{s}$ over all the vertices,
\[ \ket{\psi(0)} = \ket{s} = \frac{1}{\sqrt{N}} \sum_{i = 1}^{N} \ket{i}, \]
to a state that, when measured, collapses to a marked vertex with high probability. Note that beginning in this state expresses our initial lack of knowledge of where the marked vertices might be, so it guesses each vertex with equal probability. Furthermore, this state is an eigenstate $H_\text{walk}$, so the quantum walk alone does not change our information about where the marked vertices might be---it only changes when including the oracle, as in \eqref{eq:H}.

\begin{figure}
\begin{center}
	\subfloat[] {
		\includegraphics{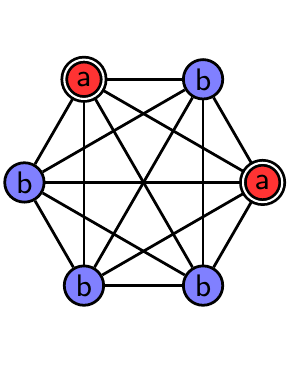}
		\label{fig:complete}
	} \quad \quad
	\subfloat[] {
		\includegraphics{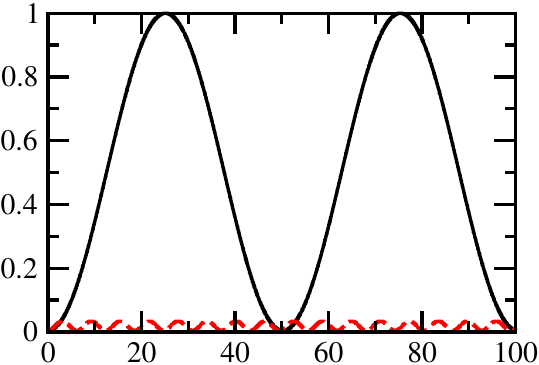}
		\label{fig:complete_evolution_a}
	}
	\caption{\textbf{(a)} The complete graph of $N = 6$ vertices, $k = 2$ of which are marked and denoted by double circles. Identically evolving vertices are identically colored and labeled. \textbf{(b)} The success probability as a function of time for search on the complete graph with $N = 1024$ and $k = 4$. The black solid curve is when $\gamma = \gamma_c = 1/N$, and the red dashed curve is when $\gamma = 2\gamma_c = 2/N$.}
\end{center}
\end{figure}

As an example, let us consider the continuous-time quantum walk formulation of Grover's algorithm \cite{Grover1996,FG1998a}. Since Grover's algorithm solves the \emph{unstructured} search problem, one can move from any vertex to any other. So this is simply search on the complete graph of $N$ vertices \cite{CG2004}, an example of which is shown in Figure~\ref{fig:complete}. 

By symmetry, the marked vertices evolve identically, as do the non-marked vertices. So we respectively group identically-evolving vertices together, as shown by identical colors and labels in Figure~\ref{fig:complete}:
\[ \ket{a} = \frac{1}{\sqrt{k}} \sum_{i \in \text{red}} \ket{i}, \quad \ket{b} = \frac{1}{\sqrt{N-k}} \sum_{i \in \text{blue}} \ket{i}. \]
Then the system evolves in a two-dimensional subspace spanned by $\{\ket{a}, \ket{b}\}$, in which the search Hamiltonian \eqref{eq:H} is
\[ H = -\gamma \begin{pmatrix}
	k-1 + \frac{1}{\gamma} & \sqrt{k(N-k)} \\
	\sqrt{k(N-k)} & N-k-1 \\
\end{pmatrix}. \]
Following the analysis of \cite{WongDissertation}, but generalized to $k$ marked vertices, this has eigenstates
\[ \ket{\psi_{0,1}} \propto \ket{s} + \frac{1 - \gamma N \pm \mathrm{\Delta} E}{2\gamma\sqrt{kN}} \ket{a} \]
with gap in the corresponding eigenvalues $E_0$ and $E_1$
\[ \mathrm{\Delta} E = E_1 - E_0 = \sqrt{(1-\gamma N)^2 + 4k\gamma}. \]
When $\gamma$ takes its critical value of $\gamma_c = 1/N$, the eigenstates are proportional to $\ket{s} \pm \ket{a}$ with an energy gap of $\mathrm{\Delta} E = 2\sqrt{k/N}$, so the system evolves from $\ket{s}$ to $\ket{a}$ in time $\pi/\mathrm{\Delta} E = (\pi/2) \sqrt{N/k}$ (see Section 3 of \cite{Wong2015b} for an explicit calculation). This is the quadratic speedup of Grover's algorithm over a classical computer's $O(N/k)$. As a check, Figure~\ref{fig:complete_evolution_a} shows the probability of measuring the quantum walker at a marked vertex as a function of time, and it reaches $1$ at time $(\pi/2) \sqrt{1024/4} = 8\pi \approx 25.133$, as expected.

From a straightforward calculation in \cite{Wong2015c} (generalized to multiple marked vertices), $\gamma$ must be chosen within $o(1/N^{3/2})$ of its critical value of $1/N$ for the algorithm to evolve from $\ket{s}$ to the marked vertices in time $(\pi/2)\sqrt{N/k}$ for large $N$. This can be relaxed to evolve to the marked vertices with constant probability in time $\Theta(\sqrt{N/k})$ if $\gamma$ is chosen within $O(1/N^{3/2})$ of its critical value of $1/N$. If $\gamma$ is further away from its critical value than this, then the initial state $\ket{s}$ converges to an eigenstate of $H$ for large $N$. So asymptotically, the system stays in $\ket{s}$ throughout its evolution, only picking up a global phase. This is shown in Figure~\ref{fig:complete_evolution_a} with $\gamma = 2\gamma_c = 2/N$; this is far enough from the critical value of $1/N$ such that the success probability plotted as a function of time converges to a flat, horizontal line for large $N$.

\begin{figure}
\begin{center}
	\subfloat[] {
		\includegraphics{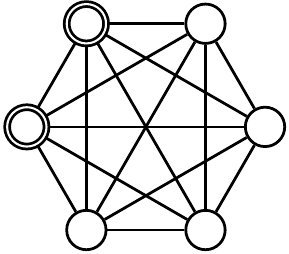}
	} \quad \quad
	\subfloat[] {
		\includegraphics{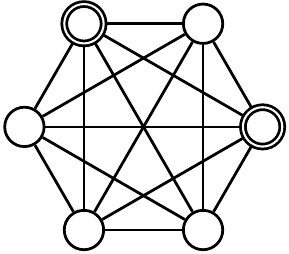}
	}
	\caption{\label{fig:complete_move} Two equivalent ways to mark $k = 2$ vertices, denoted by double circles, on the complete graph of $N = 6$ vertices.}
\end{center}
\end{figure}

Since Grover's search problem is unstructured, the problem is unchanged no matter which $k$ vertices are marked. For example, the two configurations in Figure~\ref{fig:complete_move} are equivalent (\textit{i.e.}, isomorphic). Since the location of the marked vertices does not change the structure of the problem, the algorithm (the critical jumping rate $\gamma_c = 1/N$, the evolution, the runtime of $(\pi/2)\sqrt{N/k}$, \textit{etc.}) is unchanged.

\begin{figure}
\begin{center}
	\subfloat[] {
		\includegraphics[width=2.0in]{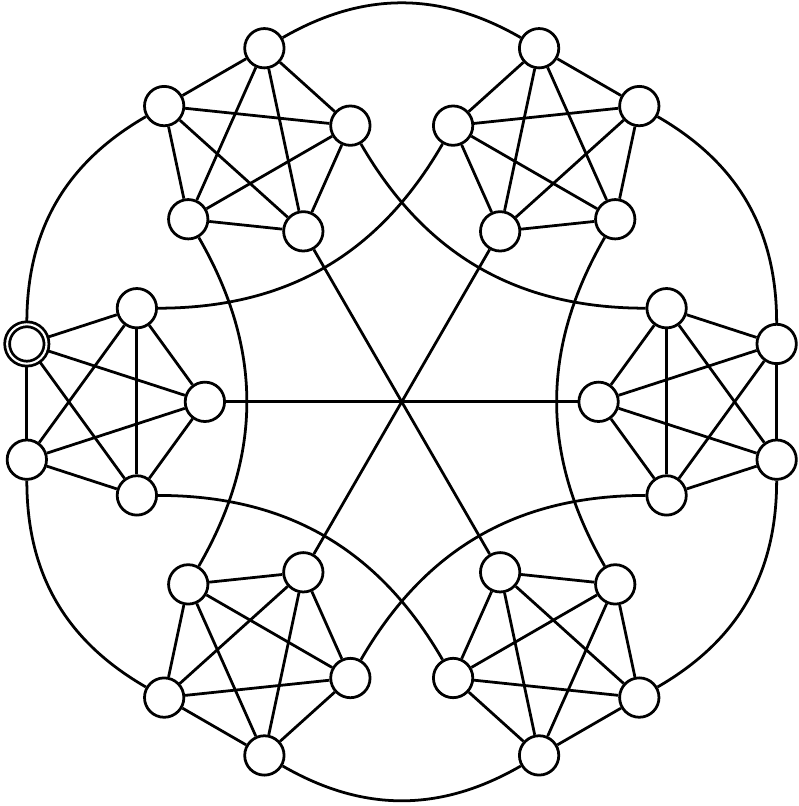}
	} \quad \quad
	\subfloat[] {
		\includegraphics[width=2.0in]{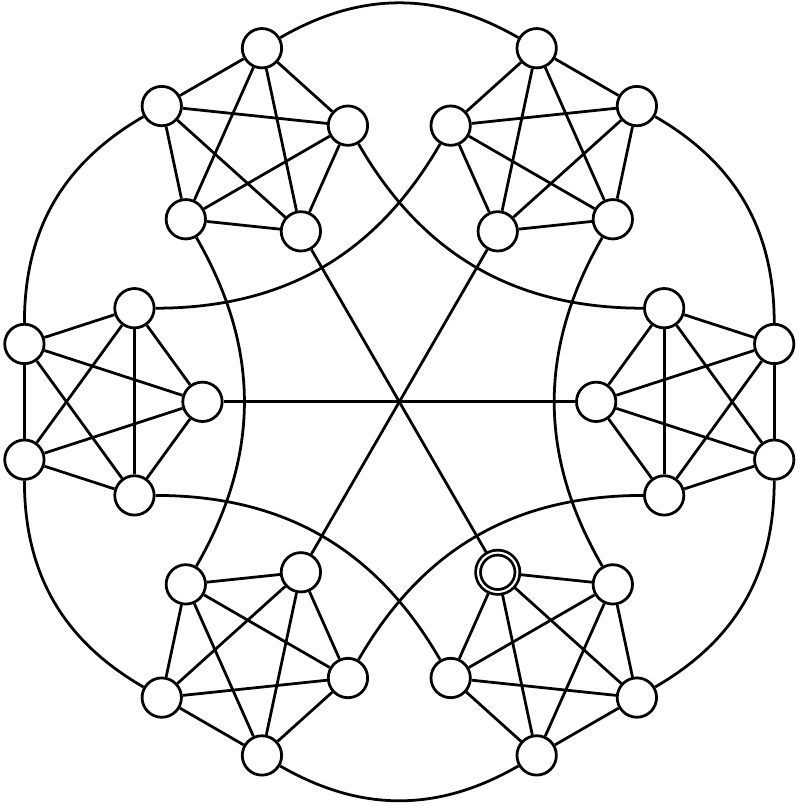}
	}
	\caption{\label{fig:simplex_one} Two equivalent ways to mark $k = 1$ vertex, denoted by double circles, on the simplex of complete graphs with $M = 5$.}
\end{center}
\end{figure}

Even for search on non-complete graphs, which is called spatial search, it is possible to retain this property that the location of the marked vertices does not matter. The most common way is to search a vertex-transitive graph for a single marked vertex. This includes the hypercube \cite{CG2004}, arbitrary-dimensional periodic square lattices \cite{CG2004}, strongly regular graphs \cite{JMW2014}, the ``simplex of complete graphs'' \cite{MeyerWong2014}, and complete bipartite graphs \cite{Novo2015}. For example, two possible ways to mark $k = 1$ vertex on the simplex of complete graphs (defined more formally later) are shown in Figure~\ref{fig:simplex_one}, and they are clearly equivalent (\textit{i.e.}, isomorphic). Another way to make the arrangement of marked vertices irrelevant is by marking a cluster of vertices, such that moving the cluster leaves the search problem unaltered \cite{WongAmbainis2015}.

\begin{figure}
\begin{center}
	\subfloat[] {
		\includegraphics[width=2.0in]{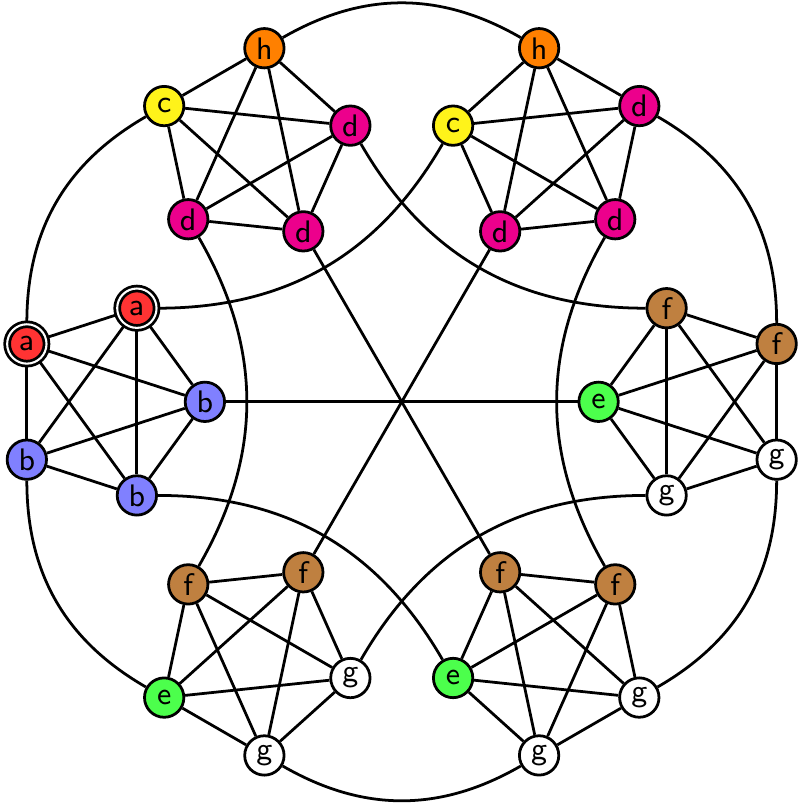}
		\label{fig:simplex_two1}
	}
	
	\subfloat[] {
		\includegraphics[width=2.0in]{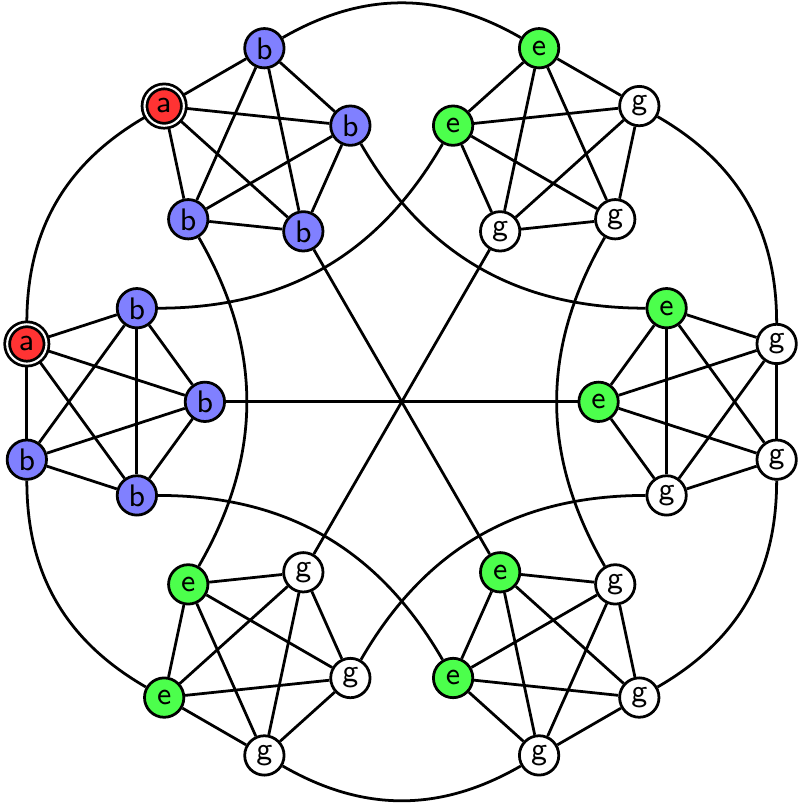}
		\label{fig:simplex_two2}
	}  \quad \quad
	\subfloat[] {
		\includegraphics[width=2.0in]{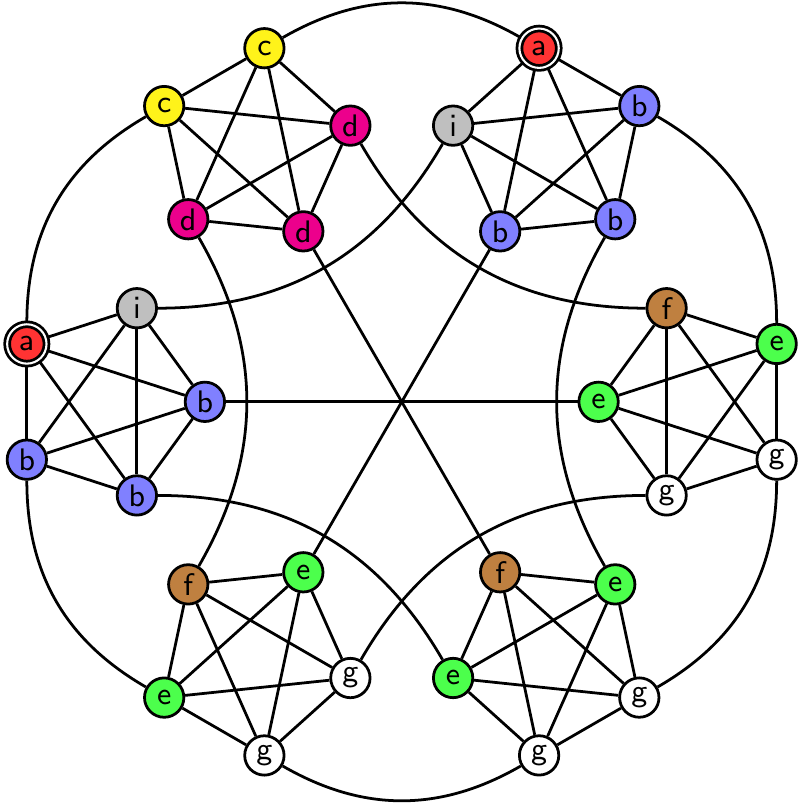}
		\label{fig:simplex_two3}
	}

	\subfloat[] {
		\includegraphics[width=2.0in]{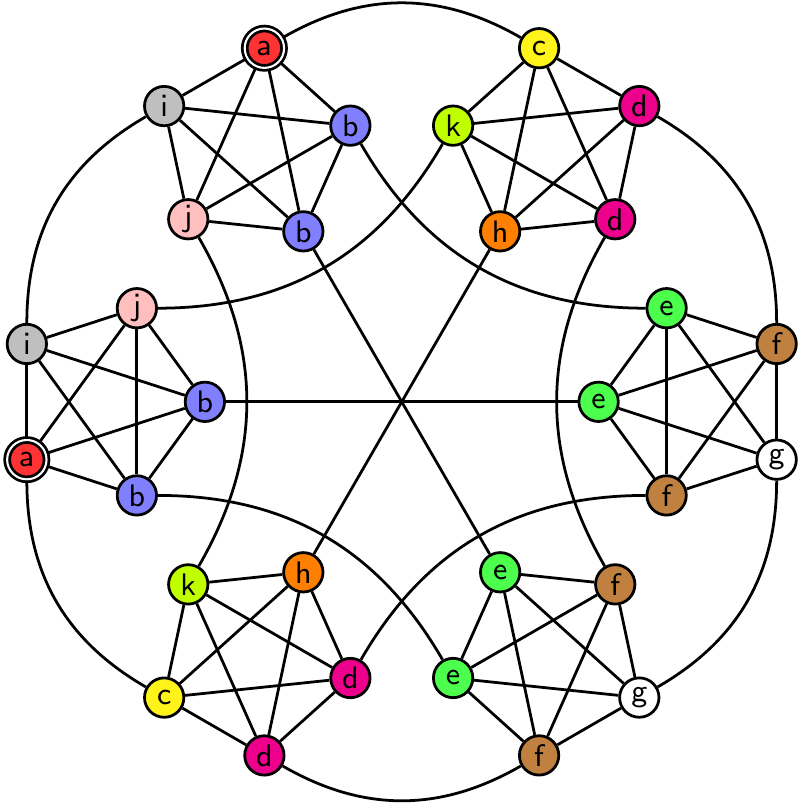}
		\label{fig:simplex_two4}
	} \quad \quad
	\subfloat[] {
		\includegraphics[width=2.0in]{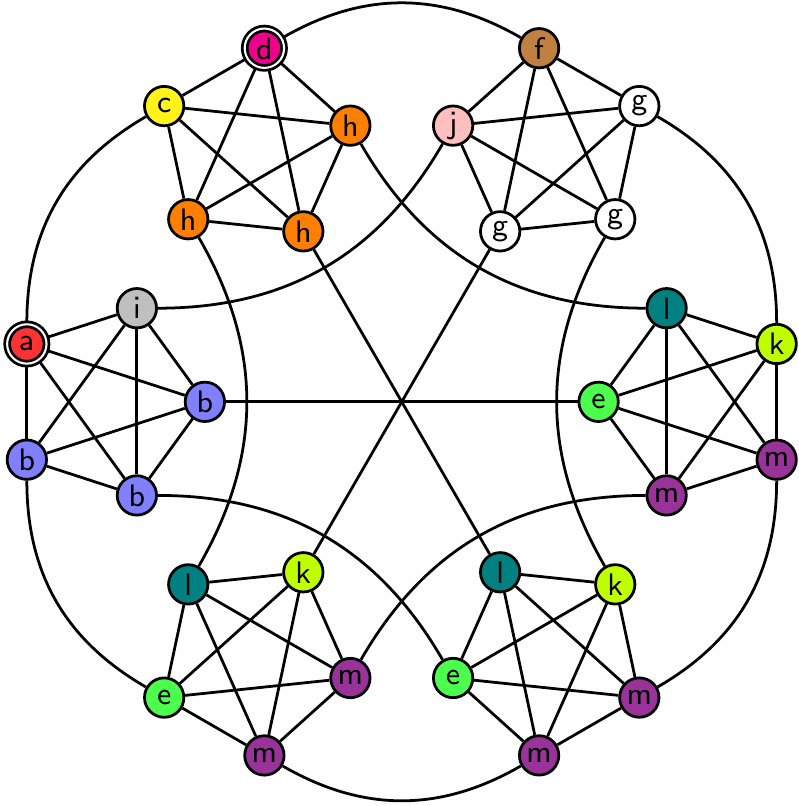}
		\label{fig:simplex_two5}
	}
	\caption{\label{fig:simplex_two}The five ways to distribute two marked vertices, indicated by double circles, on the simplex of complete graphs with $M = 5$. Identically evolving vertices are identically colored and labeled, and the labels indicate the subspace basis vectors that the vertices belong to.}
\end{center}
\end{figure}

Even though the location of a unique ($k = 1$) marked vertex on a vertex-transitive graph does not affect the search problem, when there are multiple marked vertices ($k > 1$), their locations generally do make a difference in spatial search. For example, there are five inequivalent ways to arrange $k = 2$ marked vertices on the simplex of complete graphs, and they are shown in Figure~\ref{fig:simplex_two}. The simplex of complete graphs is the $M$-simplex with each of its $M+1$ vertices replaced by a complete graph of $M$ vertices, so it has a total of $N = M(M+1)$ vertices. In this paper, we explicitly solve spatial search by continuous-time quantum walk for these five configurations, plus four more with a greater number of marked vertices. In doing so, we show that different arrangements of marked vertices can dramatically change the critical jumping rate $\gamma_c$ that is needed for the algorithm to succeed.

This shows that the algorithm is dependent on the configuration of the marked vertices. Although such a dependence has been shown for \emph{discrete}-time quantum walks \cite{Kempe2003}, this explicit demonstration seems new for \emph{continuous}-time quantum walks. This result also highlights differences between the two approaches. While $\gamma_c$ is primarily affected for continuous-time quantum walks, discrete-time quantum walks do not have this parameter. Instead, they are typically governed by a ``coin'' \cite{Meyer1996a,Meyer1996b}, and this coin is chosen in different ways to define a search problem \cite{AKR2005}. For example, with the choice from \cite{SKW2003}, one gets an algorithm that efficiently searches arbitrary-dimensional periodic square lattices with any configuration of two marked vertices \cite{AKR2005}. With more marked vertices, the runtime can change \cite{NR2015a}, even to the point of having no improvement over classically guessing \cite{AR2008}. With a different coin that yields precisely the phase flip in Grover's algorithm \cite{AKR2005,Wong2015b}, there are other exceptional configurations that cause no improvement over classical as well \cite{NR2015b}. Using Szegedy's \cite{Szegedy2004} method of defining a discrete-time quantum walk, one can also obtain a quadratic improvement in search over a classical random walk's ``extended'' hitting time when searching with multiple marked vertices \cite{KMOR2014}. This vast number of results for discrete-time quantum walks with multiple marked vertices dwarfs those for continuous-time quantum walks, of which this seems to be the first.

We choose the simplex of complete graphs for our analysis because it has enough structure to reveal interesting properties, yet enough symmetry to be analytically tractable. It was first used in quantum search to prove that connectivity is a poor indicator of fast quantum search \cite{MeyerWong2014}, and it has since been used to introduce a search algorithm that takes multiple walk-steps for each oracle query \cite{WongAmbainis2015} and to demonstrate faster search on a weighted graph \cite{Wong2015e}.

In the next section, we solve spatial search on the simplex of complete graphs with two marked vertices, whose five possible configurations were shown in Figure~\ref{fig:simplex_two}. In doing so, we show that the first case's critical jumping rate $\gamma_c$ differs from the other four. Afterwards, we solve search with a greater number of marked vertices---two cases with $M+1$ marked vertices and two cases with $2(M+1)$ marked---showing that the critical jumping rate changes in a similar manner to the $k = 2$ case. This shows that for spatial search by continuous-time quantum walk, the critical jumping rate $\gamma_c$ is dependent on the arrangement of the marked vertices.


\section{Two Marked Vertices}

The five possible configurations with $k = 2$ marked vertices were shown in Figure~\ref{fig:simplex_two}, where the marked vertices are indicated by double circles. As shown, case (a) has both marked vertices in the same complete graph, and the remaining four cases (b), (c), (d), and (e) have them in different complete graphs.

\begin{table}
\begin{center}
\caption{\label{table:simplex_two}The five cases of search with $2$ marked vertices, shown in Figure~\ref{fig:simplex_two}, with the subspace dimension and the two stage's critical jumping rate ($\gamma_c$), runtime, and evolution.}
\begin{tabular}{ccccc}
	\toprule
	Case & Dimension & $\gamma_{c}$ & Runtime & Evolution \\
	\midrule
	Figure~\ref{fig:simplex_two1} & 8D  & $\frac{3}{M} + o\left(\frac{1}{M^{5/2}}\right)$ & $\frac{\pi}{6} M^{3/2}$ & $\ket{g} \rightarrow \ket{b}$ \\[1.5ex]
	& & $\frac{1}{M} + o\left(\frac{1}{M^{3/2}}\right)$ & $\frac{\pi}{2\sqrt{2}} \sqrt{M}$ & $\ket{b} \rightarrow \ket{a}$ \\[1.5ex]
	\midrule
	Figure~\ref{fig:simplex_two2} & 4D  & $\frac{2}{M} + o\left(\frac{1}{M^{5/2}}\right)$ & $\frac{\pi}{2^{5/2}} M^{3/2}$ & $\ket{g} \rightarrow \ket{b}$ \\[1.5ex]
	& & $\frac{1}{M} + o\left(\frac{1}{M^{3/2}}\right)$ & $\frac{\pi}{2} \sqrt{M}$ & $\ket{b} \rightarrow \ket{a}$ \\[1.5ex]
	\midrule
	Figure~\ref{fig:simplex_two3} & 8D  & $\frac{2}{M} + o\left(\frac{1}{M^{5/2}}\right)$ & $\frac{\pi}{2^{5/2}} M^{3/2}$ & $\ket{g} \rightarrow \ket{b}$ \\[1.5ex]
	& & $\frac{1}{M} + o\left(\frac{1}{M^{3/2}}\right)$ & $\frac{\pi}{2} \sqrt{M}$ & $\ket{b} \rightarrow \ket{a}$ \\[1.5ex]
	\midrule
	Figure~\ref{fig:simplex_two4} & 11D  & $\frac{2}{M} + o\left(\frac{1}{M^{5/2}}\right)$ & $\frac{\pi}{2^{5/2}} M^{3/2}$ & $\ket{g} \rightarrow \ket{b}$ \\[1.5ex]
	& & $\frac{1}{M} + o\left(\frac{1}{M^{3/2}}\right)$ & $\frac{\pi}{2} \sqrt{M}$ & $\ket{b} \rightarrow \ket{a}$ \\[1.5ex]
	\midrule
	Figure~\ref{fig:simplex_two5} & 13D & $\frac{2}{M} + o\left(\frac{1}{M^{5/2}}\right)$ & $\frac{\pi}{2^{5/2}} M^{3/2}$ & $\ket{m} \rightarrow \ket{b} + \ket{h}$ \\[1.5ex]
	& & $\frac{1}{M} + o\left(\frac{1}{M^{3/2}}\right)$ & $\frac{\pi}{2} \sqrt{M}$ & $\ket{b} + \ket{h} \rightarrow \ket{a} + \ket{d}$ \\[1.5ex]
	\bottomrule
\end{tabular}
\end{center}
\end{table}

As with search with a single marked vertex \cite{MeyerWong2014}, we get two-stage algorithms for each of these five cases, where the system evolves with one critical jumping rate $\gamma_{c1}$ for some time, and then with a second critical jumping rate $\gamma_{c2}$ for a (likely different) amount of time. The detailed calculations for all five cases, including the generalization of the first case to any constant number of marked vertices in a single complete graph, are in Appendix~\ref{appendix:two}, and the main results are summarized in Table~\ref{table:simplex_two}.

\begin{figure}
\begin{center}
	\subfloat[] {
		\includegraphics[width=2.0in]{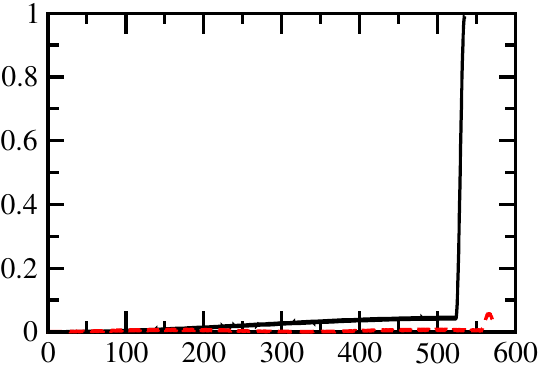}
		\label{fig:simplex_two1_evolution_a}
	} \quad \quad
	\subfloat[] {
		\includegraphics[width=2.0in]{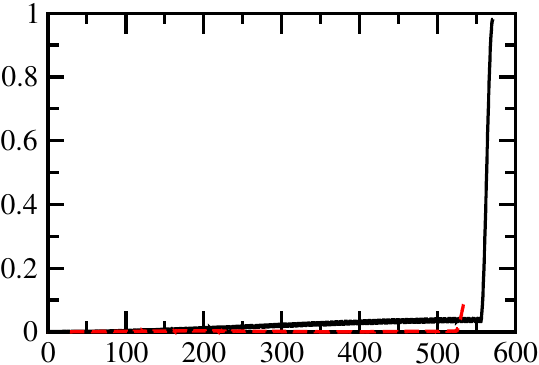}
		\label{fig:simplex_two2_evolution_a}
	}
	\caption{The success probability as a function of time for search on the simplex of complete graphs with $M = 100$ and $k = 2$ marked vertices. \textbf{(a)} Search for the first case using the correct $\gamma_c$'s and runtimes for both stages of the algorithm (black solid) and using the incorrect values from the other cases (red dashed). \textbf{(b)} Search for the second case using the correct $\gamma_c$'s and runtimes for both stages of the algorithm (black solid) and using the incorrect values from the first case (red dashed).}
\end{center}
\end{figure}

To give a sense of the calculations in Appendix~\ref{appendix:two} and explain the evolutions summarized in Table~\ref{table:simplex_two}, consider the first case in Figure~\ref{fig:simplex_two1}. The system evolves in an 8-dimensional subspace, independent of $M$, because there are only eight different kinds of vertices, as shown by the eight unique colors and labels in Figure~\ref{fig:simplex_two1}. We group identically-evolving vertices into basis vectors $\ket{a}$, $\ket{b}$, \dots, $\ket{h}$ for the 8D subspace. Then writing the search Hamiltonian \eqref{eq:H} in this 8D subspace, we find that for most values of $\gamma$, the initial state $\ket{s}$ is asymptotically an eigenvector of $H$, which means the system does not evolve except for acquiring a global, unobservable phase.

For the system to evolve, we must choose the jumping rate $\gamma$ to be within $o(1/M^{5/2})$ of its critical value $\gamma_{c1} = 3/M$ so that $\ket{s}$ experiences a ``phase transition'' \cite{CG2004}, where it becomes supported by two eigenvectors of $H$ instead of one. This can be found using degenerate perturbation theory \cite{JMW2014,Wong2014}, and it causes two eigenvectors of the Hamiltonian \eqref{eq:H} to be asymptotically proportional to $\ket{g} \pm \ket{b}$ with an energy gap of $\mathrm{\Delta} E = 6/M^{3/2}$. Since the equal superposition state $\ket{s}$ is asymptotically $\ket{g}$ for large $N$ (because the white $g$ vertices in Figure~\ref{fig:simplex_two1} overwhelming comprise most of the vertices in the graph for large $N$), near $\gamma = \gamma_c$, the system evolves from $\ket{s}$ to $\ket{b}$ in time $t_1 = \pi/\mathrm{\Delta} E = \pi M^{3/2} / 6$. From Figure~\ref{fig:simplex_two1}, this means probability has now collected at the correct complete graph (\textit{i.e.}, at the blue $b$ vertices), but is not yet at the marked red $a$ vertices within that complete graph. This is summarized in the first row of information in Table~\ref{table:simplex_two}.

For the second stage of the algorithm, we want to move the probability from $\ket{b}$ to the marked vertices $\ket{a}$. As shown in Appendix~\ref{appendix:two}, the Hamiltonian has two eigenstates that are asymptotically proportional to $\ket{b} \pm \ket{a}$ when the jumping rate $\gamma$ takes the critical value $\gamma_{c2} = 1/M$ with an energy gap of $\mathrm{\Delta} E = 2^{3/2}/\sqrt{M}$. So the system now evolves from $\ket{b}$ to $\ket{a}$ in time $t_2 = \pi / \mathrm{\Delta} E = \pi \sqrt{M}/2^{3/2}$. This causes the system to evolve from the blue $b$ vertices to the red $a$ vertices, which are marked, accomplishing the search. This is summarized in the second row of information in Table~\ref{table:simplex_two}.

The success probability for the entire evolution of the algorithm is shown in Figure~\ref{fig:simplex_two1_evolution_a} as the solid black curve. For most of the time, the system is in the first stage of the evolution, where probability asymptotically builds up at $\ket{b}$. Any buildup in $\ket{a}$ during this time is due to higher-order contributions that are negligible for large $N$. Then in the second stage of the algorithm, the probability quickly shifts from $\ket{b}$ to $\ket{a}$, indicated by the sudden spike in success probability in Figure~\ref{fig:simplex_two1_evolution_a}.

Examining search for all five configurations in Table~\ref{table:simplex_two} reveals that the first configuration behaves differently from the other four---it has a different critical jumping rate for the first stage of the algorithm $\gamma_{c1}$, and the runtime of both stages is different. Of these, the important difference is the critical jumping rate, which truly is different because $3/M$ and $2/M$ are separated by a gap of more than $o(1/M^{5/2})$. Thus evolving by the wrong configuration's value would cause the system to asymptotically stay in its initial state $\ket{s}$ throughout its evolution, only acquiring a global, unobservable phase. This is shown in Figure~\ref{fig:simplex_two1_evolution_a}, where the dashed red curve is search for the first configuration, but incorrectly using the critical $\gamma$'s and runtimes from the other cases. Note that the probability stays small, and any buildup from higher-order corrections vanishes for large $N$. Similarly, Figure~\ref{fig:simplex_two2_evolution_a} shows search with vertices in the second configuration in Figure~\ref{fig:simplex_two2}---with the correct critical $\gamma$'s and runtimes, success probability builds up as desired, but with the wrong values from the first configuration, it fails to build up. Note that the second, third, fourth, and fifth configurations corresponding to Figures~\ref{fig:simplex_two2}, \ref{fig:simplex_two3}, \ref{fig:simplex_two4}, and \ref{fig:simplex_two5}, have identical critical jumping rates and runtimes. So while rearranging marked vertices can change the critical jumping rate substantially, it could also do nothing.

Given the extensive calculations in Appendix~\ref{appendix:two} that are needed to derive the results in Table~\ref{table:simplex_two}, manually working through all configurations of $k$ marked vertices on the simplex of complete graphs is impractical using the current method. We do analyze in the next section, however, four additional configurations with a greater number of marked vertices, two of which show that moving marked vertices again changes the critical jumping rate, and two that show no change.


\section{Larger Examples}

\begin{figure}
\begin{center}
	\subfloat[] {
		\includegraphics[width=2.0in]{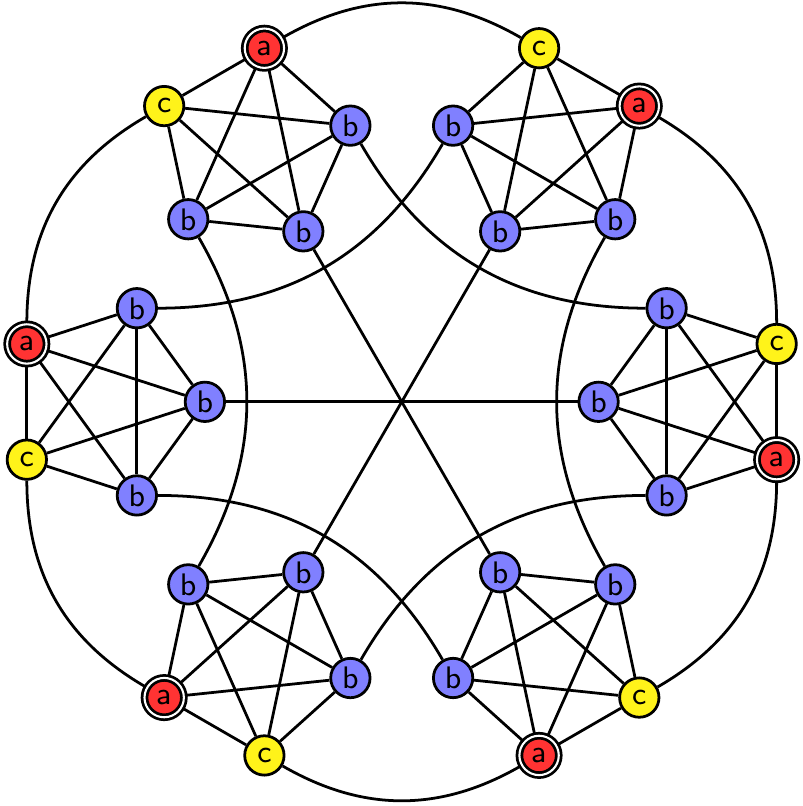}
		\label{fig:simplex_M1a}
	} \quad \quad
	\subfloat[] {
		\includegraphics[width=2.0in]{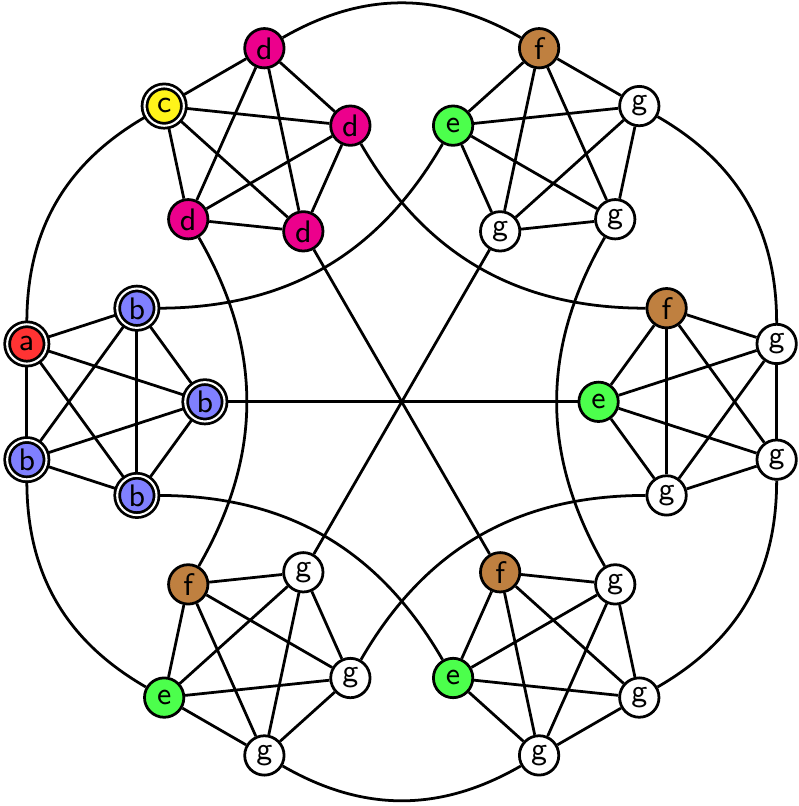}
		\label{fig:simplex_M1b}
	}
	\caption{\label{fig:simplex_M1}Two ways to distribute $M+1$ marked vertices, indicated double circles, on the simplex of complete graphs with $M = 5$. Identically evolving vertices are identically colored and labeled, and the labels indicate the subspace basis vectors that the vertices belong to.}
\end{center}
\end{figure}

\begin{figure}
\begin{center}
	\subfloat[] {
		\includegraphics[width=2.0in]{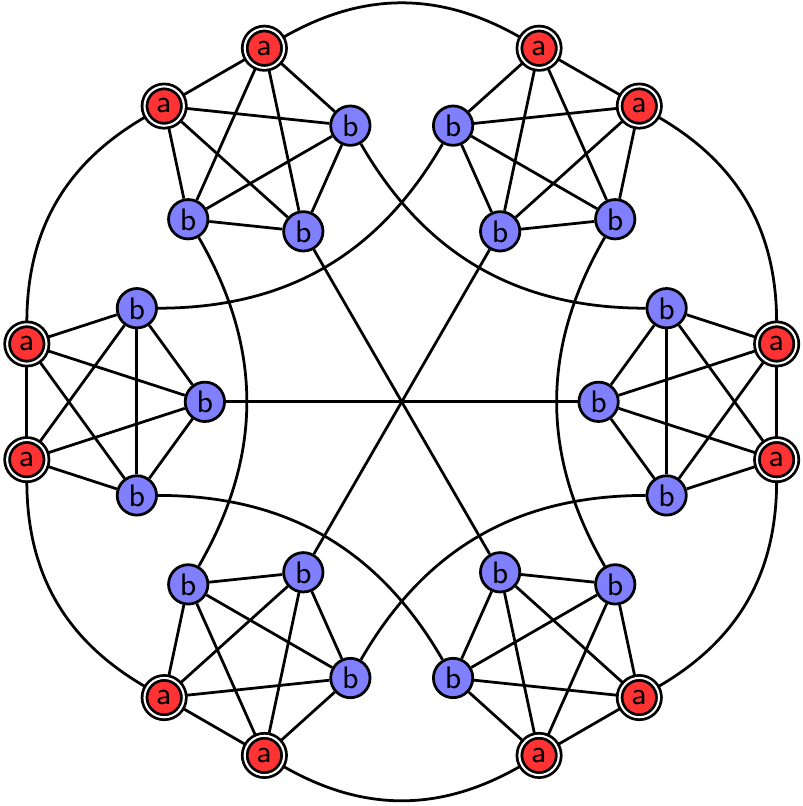}
		\label{fig:simplex_2M1a}
	} \quad \quad
	\subfloat[] {
		\includegraphics[width=2.0in]{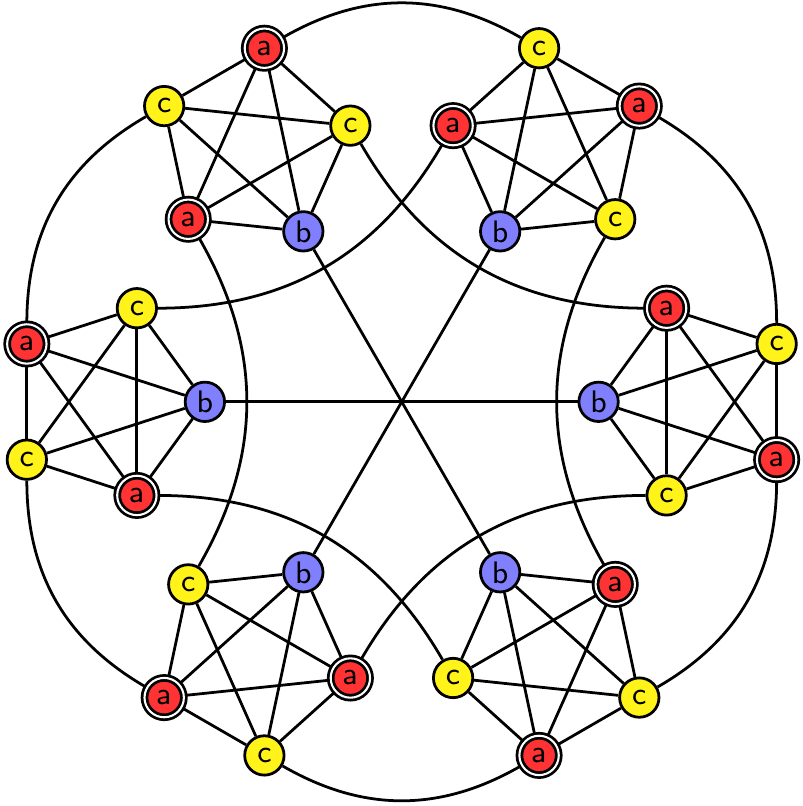}
		\label{fig:simplex_2M1b}
	}
	\caption{\label{fig:simplex_2M1}Two ways to distribute $2(M+1)$ marked vertices, indicated double circles, on the simplex of complete graphs with $M = 5$. Identically evolving vertices are identically colored and labeled, and the labels indicate the subspace basis vectors that the vertices belong to.}
\end{center}
\end{figure}

Let us consider four examples with a larger number of marked vertices. The first two are shown in Figure~\ref{fig:simplex_M1}, and they are just two of the many ways to arrange $M+1$ marked vertices on the simplex of complete graphs; in subfigure (a), they are distributed so that each complete graph has one marked vertex, and in subfigure (b), all the marked vertices are clustered in a single complete graph, except for one. The other two configurations are shown in Figure~\ref{fig:simplex_2M1}, and they are just two of the many ways to distribute $2(M+1)$ marked vertices so that each complete graph has two marked vertices.

\begin{table}
\begin{center}
	\caption{\label{table:simplex_larger}Some cases of search with multiple marked vertices with the subspace dimension, critical jumping rate ($\gamma_c$), runtime, and evolution.}
\begin{tabular}{ccccc}
	\toprule
	Case & Dimension & $\gamma_c$ & Runtime & Evolution \\
	\midrule
	Figure~\ref{fig:simplex_M1a} & 3D & $\frac{1}{M} + o\left(\frac{1}{M^{3/2}}\right)$ & $\frac{\pi}{2} \sqrt{M}$ & $\ket{b} \rightarrow \ket{a}$ \\[1.5ex]
	Figure~\ref{fig:simplex_M1b} & 7D & $1 + \frac{3}{M} + o\left(\frac{1}{M^{3/2}}\right)$ & $\frac{\pi}{2} \sqrt{M}$ & $\ket{g} \rightarrow \ket{b}$ \\[0.8ex]
	\midrule
	Figure~\ref{fig:simplex_2M1a} & 2D & $\frac{1}{M} + o\left(\frac{1}{M^{3/2}}\right)$ & $\frac{\pi}{2\sqrt{2}} \sqrt{M}$ & $\ket{b} \rightarrow \ket{a}$ \\[1.1ex]
	Figure~\ref{fig:simplex_2M1b} & 3D & $\frac{1}{M} + o\left(\frac{1}{M^{3/2}}\right)$ & $\frac{\pi}{2\sqrt{2}} \sqrt{M}$ & $\ket{b} \rightarrow \ket{a}$ \\
	\bottomrule
\end{tabular}
\end{center}
\end{table}

Unlike search with a single marked vertex \cite{MeyerWong2014} or two marked vertices in the previous section, search on these graphs are single-stage algorithms. The detailed proofs are given in Appendix~\ref{appendix:larger}, and they use the same techniques as the configurations with two marked vertices. The results are summarized in Table~\ref{table:simplex_larger}, and they show that the critical jumping rate $\gamma_c$ for the two configurations in Figure~\ref{fig:simplex_M1} differ substantially enough that the search algorithm will fail (\textit{i.e.}, the system will asymptotically stay in its initial state) if the wrong $\gamma_c$ is used. In Figure~\ref{fig:simplex_2M1}, however, the critical jumping rate $\gamma_c$ and runtime are the same. So again, we see that rearranging the marked vertices can change the critical jumping rate substantially, but it also might not change it at all.


\section{Conclusion}

We have shown that when there are multiple marked vertices, their configuration on a graph can affect the critical jumping rate $\gamma_c$ of the continuous-time quantum walk. Previous work on search by continuous-time quantum walk has avoided this effect by restricting search to a single marked vertex on vertex-transitive graphs, or by clustering marked vertices together. This highlights a difference between continuous- and discrete-time quantum walks as they are used for search, since discrete-time quantum walks do not have the jumping rate as a parameter.

This work leaves open how general configurations of multiple marked vertices affect search on the simplex of complete graphs, since our analysis only examined nine specific arrangements. As a curious observation, our results follow a pattern: the critical jumping rate $\gamma_c$ is affected by the number of marked vertices in each complete graph, not by their arrangement within them. That is, going from Figure~\ref{fig:simplex_M1}a to Figure~\ref{fig:simplex_M1}b moves marked vertices \emph{across} complete graphs, changing the number within each, which changes $\gamma_c$. On the other hand, going from Figure~\ref{fig:simplex_2M1}a to Figure~\ref{fig:simplex_2M1}b moves marked vertices \emph{within} complete graphs, and this rearrangement within complete graphs makes no difference to $\gamma_c$. This behavior is consistent with two marked vertices in Figure~\ref{fig:simplex_two}, and it might stem from each complete graph being sufficiently connected within itself that the arrangement of marked vertices within them does not matter. Whether this pattern holds in general is a subject of further investigation, as is how continuous-time quantum walks search other graphs with multiple marked vertices.


\begin{acknowledgements}
	Thanks to Andris Ambainis for useful discussions. This work was supported by the European Union Seventh Framework Programme (FP7/2007-2013) under the QALGO (Grant Agreement No.~600700) project, and the ERC Advanced Grant MQC. 
\end{acknowledgements}


\bibliographystyle{qinp}
\bibliography{refs}


\appendix
\section{\label{appendix:two}Details for Two Marked Vertices}

In this appendix, we employ degenerate perturbation theory \cite{Sakurai1993,JMW2014,Wong2014} to find the critical $\gamma$'s and runtimes for search with two marked vertices, of which there are five cases, as summarized in Table~\ref{table:simplex_two}.

\subsection{Two Marked, Case 1, Generalized to Constant Marked Vertices}

	Instead of having just $2$ marked vertices in a single complete graph, we generalize the problem to $k$ constant marked vertices. Even with this generalization, the system still evolves in an 8-dimensional subspace, as shown in Figure~\ref{fig:simplex_two1}, spanned by
	\begin{align*}
		&\ket{a} = \frac{1}{\sqrt{k}} \sum_{i \in \text{red}} \ket{i}, && \ket{e} = \frac{1}{\sqrt{M-k}} \sum_{i \in \text{green}} \ket{i}, \\
		&\ket{b} = \frac{1}{\sqrt{M-k}} \sum_{i \in \text{blue}} \ket{i}, && \ket{f} = \frac{1}{\sqrt{k(M-k)}} \sum_{i \in \text{brown}} \ket{i}, \\
		&\ket{c} = \frac{1}{\sqrt{k}} \sum_{i \in \text{yellow}} \ket{i}, && \ket{g} = \frac{1}{\sqrt{(M-k-1)(M-k)}} \sum_{i \in \text{white}} \ket{i}, \\
		&\ket{d} = \frac{1}{\sqrt{k(M-k)}} \sum_{i \in \text{magenta}} \ket{i}, && \ket{h} = \frac{1}{\sqrt{k(k-1)}} \sum_{i \in \text{orange}} \ket{i}. 
	\end{align*}
	In this subspace, the search Hamiltonian \eqref{eq:H} is
	\[ \setlength{\arraycolsep}{1pt} H = -\gamma \begin{pmatrix}
		k-1 + \frac{1}{\gamma} & \sqrt{k M_k} & 1 & 0 & 0 & 0 & 0 & 0 \\
		\sqrt{k M_k} & M_{k1} & 0 & 0 & 1 & 0 & 0 & 0 \\
		1 & 0 & 0 & \sqrt{M_k} & 0 & 0 & 0 & \sqrt{k-1} \\
		0 & 0 & \sqrt{M_k} & M_{k1} & 0 & 1 & 0 & \sqrt{M_k(k-1)} \\
		0 & 1 & 0 & 0 & 0 & \sqrt{k} & \sqrt{M_{k1}} & 0 \\
		0 & 0 & 0 & 1 & \sqrt{k} & k-1 & \sqrt{k M_{k1}} & 0 \\
		0 & 0 & 0 & 0 & \sqrt{M_{k1}} & \sqrt{k M_{k1}} & M_{k1} & 0 \\
		0 & 0 & \sqrt{k-1} & \sqrt{M_k(k-1)} & 0 & 0 & 0 & k-1 \\
	\end{pmatrix}, \]
	where $M_k = M-k$ and $M_{k1} = M - k - 1$.

	\begin{figure}
	\begin{center}
		\subfloat[] {
			\includegraphics[scale=0.7]{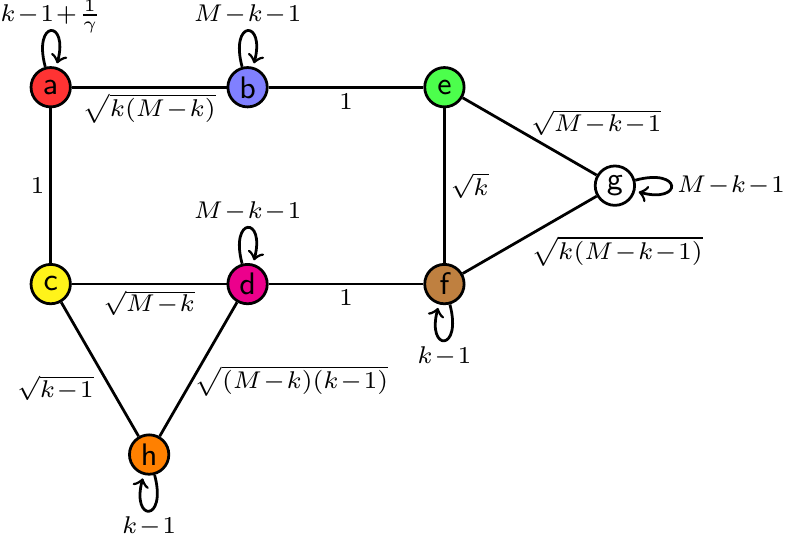}
		}

		\subfloat[] {
			\includegraphics[scale=0.7]{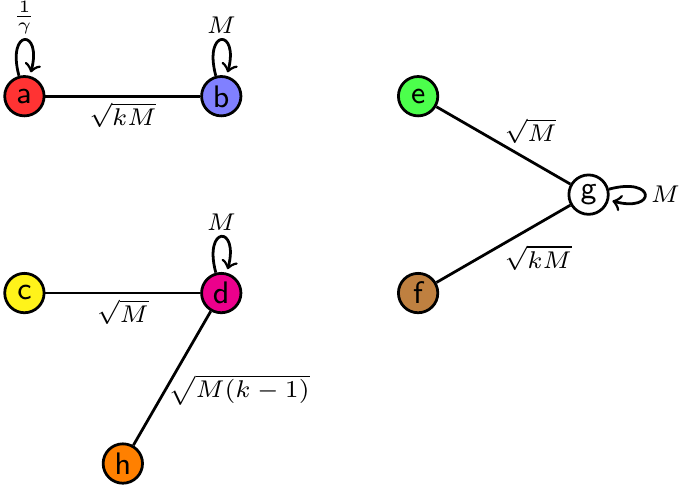}
		} \quad \quad
		\subfloat[] {
			\includegraphics[scale=0.7]{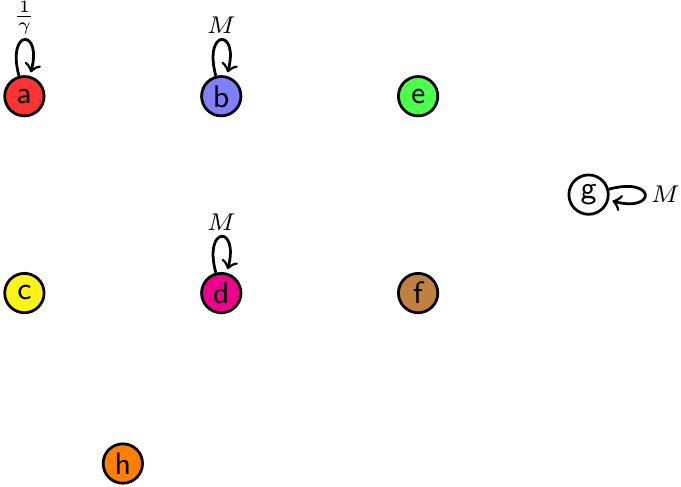}
		}
		\caption{\label{fig:simplex_two1_diagrams}Apart from a factor of $-\gamma$, \textbf{(a)} the Hamiltonian for the first case of search on the simplex of complete graphs with $k = 2$ marked vertices, \textbf{(b)} the leading-order terms for the first stage of the algorithm, and \textbf{(c)} the leading-order terms for the second stage of the algorithm. }
	\end{center}
	\end{figure}

	Using the diagrammatic approach in \cite{Wong2014} as a guide, this Hamiltonian can be visualized as a graph with eight vertices, as shown in Figure~\ref{fig:simplex_two1_diagrams}a. For the first stage of the algorithm, the leading-order Hamiltonian $H^{(0)}$ can be visualized as shown in Figure~\ref{fig:simplex_two1_diagrams}b, where we have excluded edges that scale less than $\sqrt{M}$. From this, the eight eigenvectors of $H^{(0)}$ are easily seen:~two are linear combinations of $\ket{a}$ and $\ket{b}$, three are linear combinations of $\ket{c}$, $\ket{d}$, and $\ket{h}$, and the final three are linear combinations of $\ket{e}$, $\ket{f}$, and $\ket{g}$. They correspond to the eigenvectors of
	\begin{gather*}
		H_{ab}^{(0)} = -\gamma \begin{pmatrix}
			\frac{1}{\gamma} & \sqrt{kM} \\
			\sqrt{kM} & M \\
		\end{pmatrix}, \\
		H_{cdh}^{(0)} = -\gamma \begin{pmatrix}
			0 & \sqrt{M} & 0 \\
			\sqrt{M} & M & \sqrt{M(k-1)} \\
			0 & \sqrt{M(k-1)} & 0 \\
		\end{pmatrix}, \\
		H_{efg}^{(0)} = -\gamma \begin{pmatrix}
			0 & 0 & \sqrt{M} \\
			0 & 0 & \sqrt{kM} \\
			\sqrt{M} & \sqrt{kM} & M \\
		\end{pmatrix}.
	\end{gather*}
	Since $\ket{s} \approx \ket{g}$, and we want probability to move towards the marked vertices $\ket{a}$, we want to choose $\gamma$ so that a linear combination of $\ket{e}, \ket{f}$, and $\ket{g}$ is degenerate with a linear combination of $\ket{a}$ and $\ket{b}$. In particular, the eigenstates that we want to be degenerate are
	\[ u = \frac{2}{\sqrt{M}+\sqrt{4+4 k+M}} \ket{e} + \frac{2 \sqrt{k M}}{\sqrt{M} \left(\sqrt{M}+\sqrt{4+4 k+M}\right)} \ket{f} + \ket{g} \]
	with corresponding eigenvalue
	\[ E_u = \frac{-\gamma}{2} \left(M +\sqrt{M} \sqrt{4+4 k+M} \right) \]
	and
	\[ v = \frac{1-M \gamma +\sqrt{1-2 M \gamma +4 k M \gamma ^2+M^2 \gamma ^2}}{2 \sqrt{k M} \gamma } \ket{a} + \ket{b} \]
	with corresponding eigenvalue
	\[ E_v = \frac{1}{2} \left(-1-M \gamma -\sqrt{1-2 M \gamma +4 k M \gamma ^2+M^2 \gamma ^2}\right). \]
	Written this way, $u$ and $v$ are unnormalized, whereas $\ket{u}$ and $\ket{v}$ are their normalized versions. These eigenstates are degenerate when $\gamma$ takes its critical value of
	\[ \gamma_{c1} = \frac{-M+\sqrt{M} \sqrt{4+4 k+M}}{2 M} \approx \frac{1+k}{M}. \]
	The perturbation $H^{(1)}$, which restores terms of constant weight, causes certain linear combinations
	\[ \alpha_u \ket{u} + \alpha_v \ket{v} \]
	of these states to be eigenstates of $H^{(0)} + H^{(1)}$ \cite{Sakurai1993,JMW2014}. The coefficients $\alpha_u$ and $\alpha_v$ can be found by solving
	\[ \begin{pmatrix}
		H_{uu} & H_{uv} \\
		H_{vu} & H_{vv} \\
	\end{pmatrix} \begin{pmatrix}
		\alpha_u \\
		\alpha_v \\
	\end{pmatrix} = E \begin{pmatrix}
		\alpha_u \\
		\alpha_v \\
	\end{pmatrix}, \]
	where $H_{uv} = \langle u | H^{(0)} + H^{(1)} | v \rangle$, \textit{etc}. Solving this, the perturbed eigenvectors for large $N$ with their corresponding eigenvalues are
	\begin{gather*}
		\frac{1}{\sqrt{2}} \left( \ket{u} + \ket{v} \right), \quad E = -(k+1) + \frac{k^2+2k+1}{M} - \frac{k+1}{M^{3/2}} \\
		\frac{1}{\sqrt{2}} \left( \ket{u} - \ket{v} \right), \quad E = -(k+1) + \frac{k^2+2k+1}{M} + \frac{k+1}{M^{3/2}}
	\end{gather*}
	Since $\ket{u} \approx \ket{g}$ and $\ket{v} \approx \ket{b}$ for large $N$, the system evolves from $\ket{s} \approx \ket{g}$ to $\ket{b}$ in time $\pi/\mathrm{\Delta} E$, which is
	\[ t_1 = \frac{\pi M^{3/2}}{2(k+1)}. \]
	Diagrammatically, the perturbation $H^{(1)}$ restores edges of constant weight in Figure~\ref{fig:simplex_two1_diagrams}b, and probability flows between $\ket{g}$ and $\ket{b}$ since they are the most dominant terms.

	Using the approach of Section VI of \cite{Wong2015e}, if $\gamma$ is within $\epsilon$ of its critical value of $\gamma_{c1} \approx (1+k)/M$, then the eigenvalues of $\ket{u}$ and $\ket{v}$ now include leading-order (in $\epsilon$) terms $-\epsilon M$. In the perturbative calculation, this introduces terms scaling as $\epsilon M$ due to $H_{uu}$ and $H_{vv}$, so for this to not influence the energy gap $\Theta(1/M^{3/2})$, we must have $\epsilon M = o(1/M^{3/2})$, or $\epsilon = o(1/M^{5/2})$. Thus for the first stage of the algorithm to asymptotically evolve from $\ket{s}$ to $\ket{b}$, we require $\gamma = \gamma_{c1} + o(1/M^{5/2})$. Note if we relax this to evolve to $\ket{b}$ with constant probability, then $\gamma = \gamma_{c1} + O(1/M^{5/2})$ suffices.

	For the second stage of the algorithm, we take the leading-order Hamiltonian $H^{(0)}$ to only include edges of weight $\Theta(M)$, and its diagram is shown in Figure~\ref{fig:simplex_two1_diagrams}c. From this, the eight eigenvectors of $H^{(0)}$ are simply the basis vectors $\ket{a}$, $\ket{b}$, \dots, $\ket{h}$ with corresponding eigenvalues $-1$, $-\gamma M$, \dots, $0$. When $\gamma$ takes its critical value of
	\[ \gamma_{c2} = \frac{1}{M}, \]
	the eigenstates $\ket{a}$, $\ket{b}$, $\ket{d}$, and $\ket{g}$ of $H^{(0)}$ are degenerate with eigenvalue $-1$. Then the perturbation $H^{(1)}$, which restores terms $\Theta(\sqrt{M})$, causes certain linear combinations
	\[ \alpha_a \ket{a} + \alpha_b \ket{b} + \alpha_d \ket{d} + \alpha_g \ket{g} \]
	of these states to be eigenstates of $H^{(0)} + H^{(1)}$. The coefficients $\alpha_a$, $\alpha_b$, $\alpha_d$, and $\alpha_g$ can be found by solving
	\[ \begin{pmatrix}
		H_{aa} & H_{ab} & H_{ad} & H_{ag} \\
		H_{ba} & H_{bb} & H_{bd} & H_{bg} \\
		H_{da} & H_{db} & H_{dd} & H_{dg} \\
		H_{ga} & H_{gb} & H_{gd} & H_{gg} \\
	\end{pmatrix} \begin{pmatrix}
		\alpha_a \\
		\alpha_b \\
		\alpha_d \\
		\alpha_g \\
	\end{pmatrix} = E \begin{pmatrix}
		\alpha_a \\
		\alpha_b \\
		\alpha_d \\
		\alpha_g \\
	\end{pmatrix}, \]
	where $H_{ab} = \langle a | H^{(0)} + H^{(1)} | b \rangle$, \textit{etc}. Solving this, the perturbed eigenvectors with their corresponding eigenvalues are
	\begin{gather*}
		\ket{\psi_0} = \frac{1}{\sqrt{2}} ( \ket{b} + \ket{a} ), \quad E_0 = -1 - \sqrt{\frac{k}{M}} \\
		\ket{\psi_1} = \ket{d}, \quad E_1 = -1 \\
		\ket{\psi_2} = \ket{g}, \quad E_2 = -1 \\
		\ket{\psi_3} = \frac{1}{\sqrt{2}} ( \ket{b} - \ket{a} ), \quad E_3 = -1 + \sqrt{\frac{k}{M}}
	\end{gather*}
	So the system evolves from $\ket{b}$ to $\ket{a}$ in time $\pi / \mathrm{\Delta} E$:
	\[ t_2 = \frac{\pi}{2} \sqrt{\frac{M}{k}}. \]
	Diagrammatically, the perturbation $H^{(1)}$ restores edges $\Theta(\sqrt{M})$ in Figure~\ref{fig:simplex_two2_diagrams}c, and probability flows between $\ket{b}$ and $\ket{a}$.

	We can again use the method of \cite{Wong2015e} to find how precisely $\gamma$ must be chosen to its critical value $\gamma_{c2} = 1/M$---a straightforward calculation shows that it must be within $o(1/M^{3/2})$.

	For $k = 2$ marked vertices, the critical $\gamma$'s and runtimes are
	\[ \gamma_{c1} = \frac{3}{M}, \quad t_1 = \frac{\pi M^{3/2}}{6}, \quad \gamma_{c2} = \frac{1}{M}, \quad t_2 = \frac{\pi}{2} \sqrt{\frac{M}{2}}, \]
	all of which are in agreement with Table~\ref{table:simplex_two}.

\subsection{Two Marked, Case 2}

	As shown in Figure~\ref{fig:simplex_two2}, the system evolves in a 4-dimensional subspace spanned by
	\begin{align*}
		&\ket{a} = \frac{1}{\sqrt{2}} \sum_{i \in \text{red}} \ket{i}, && \ket{e} = \frac{1}{\sqrt{2(M-1)}} \sum_{i \in \text{green}} \ket{i},\\
	       	&\ket{b} = \frac{1}{\sqrt{2(M-1)}} \sum_{i \in \text{blue}} \ket{i}, && \ket{g} = \frac{1}{\sqrt{(M-1)(M-2)}} \sum_{i \in \text{white}} \ket{i},
	\end{align*}
	where the labels have been chosen this way to match the behavior of the vertices in the first case in Figure~\ref{fig:simplex_two1}. In this subspace, the search Hamiltonian \eqref{eq:H} is
	\[ H = -\gamma \begin{pmatrix}
		1 + \frac{1}{\gamma} & \sqrt{M-1} & 0 & 0 \\
		\sqrt{M-1} & M-2 & 1 & 0 \\
		0 & 1 & 1 & \sqrt{2(M-2)} \\
		0 & 0 & \sqrt{2(M-2)} & M-2 \\
	\end{pmatrix}. \]

	\begin{figure}
	\begin{center}
		\subfloat[] {
			\includegraphics[scale=0.7]{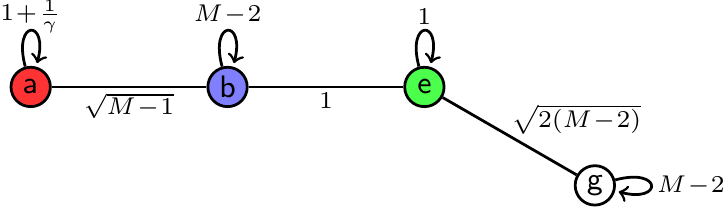}
		}

		\subfloat[] {
			\includegraphics[scale=0.7]{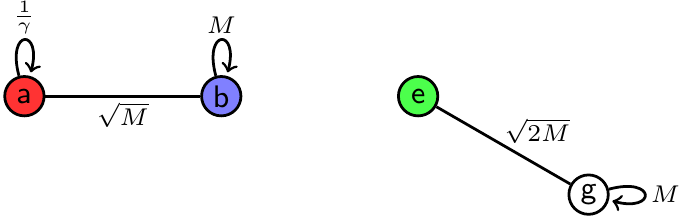}
		} \quad \quad
		\subfloat[] {
			\includegraphics[scale=0.7]{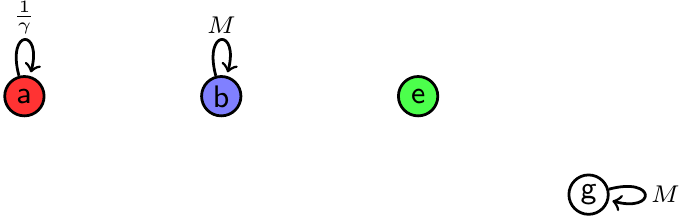}
		}
		\caption{\label{fig:simplex_two2_diagrams}Apart from a factor of $-\gamma$, \textbf{(a)} the Hamiltonian for the second case of search on the simplex of complete graphs with $k = 2$ marked vertices, \textbf{(b)} the leading-order terms for the first stage of the algorithm, and \textbf{(c)} the leading-order terms for the second stage of the algorithm. }
	\end{center}
	\end{figure}

	This Hamiltonian can be visualized as shown in Figure~\ref{fig:simplex_two2_diagrams}a. For the first stage of the algorithm, the leading-order Hamiltonian $H^{(0)}$ excludes edges that scale less than $\sqrt{M}$, and it can be visualized as shown in Figure~\ref{fig:simplex_two2_diagrams}b. The two eigenstates of $H^{(0)}$ that we want to be degenerate are
	\[ u = -\frac{\sqrt{M}-\sqrt{M+8}}{2 \sqrt{2}} \ket{e} + \ket{g} \]
	with corresponding eigenvalue
	\[ E_u = \frac{-\gamma}{2} \left(M + \sqrt{M} \sqrt{M+8} \right) \]
	and
	\[ v = -\frac{-1+M \gamma -\sqrt{1-2 M \gamma +4 M \gamma ^2+M^2 \gamma ^2}}{2 \sqrt{M} \gamma } \ket{a} + \ket{b} \]
	with corresponding eigenvalue
	\[ E_v = \frac{1}{2} \left(-1-M \gamma -\sqrt{1-2 M \gamma +4 M \gamma ^2+M^2 \gamma ^2}\right). \]
	These are degenerate when $\gamma$ takes its critical value of
	\[ \gamma_{c1} = \frac{-M+\sqrt{M} \sqrt{M+8}}{2 M} = \frac{2}{M} - \frac{4}{M^2} + O(1/M^3). \]
	The perturbation $H^{(1)}$, which restores terms of constant weight, causes certain linear combinations $\alpha_u \ket{u} + \alpha_v \ket{v}$ to be eigenstates of $H^{(0)} + H^{(1)}$. Doing the perturbative calculation to find the coefficients (as in the first case), the perturbed eigenstates for large $N$ are
	\begin{gather*}
		\frac{1}{\sqrt{2}} ( \ket{u} + \ket{v} ), \quad E = -2 - \frac{2\sqrt{2}}{M^{3/2}} \\
		\frac{1}{\sqrt{2}} ( \ket{u} - \ket{v} ), \quad E = -2 + \frac{2\sqrt{2}}{M^{3/2}}.
	\end{gather*}
	Since $\ket{u} \approx \ket{g}$ and $\ket{v} \approx \ket{b}$ for large $N$, the system evolves from $\ket{s} \approx \ket{g}$ to $\ket{b}$ in time $\pi / \mathrm{\Delta} E$:
	\[ t_1 = \frac{\pi M^{3/2}}{4\sqrt{2}}. \]
	Diagrammatically, the perturbation $H^{(1)}$ restores edges of constant weight in Figure~\ref{fig:simplex_two2_diagrams}b, and probability flows between $\ket{g}$ and $\ket{b}$ since they are the most dominant terms.

	As in the first case in the precious section, we can use the method of \cite{Wong2015e} to find how precisely $\gamma$ must be chosen to its critical value $\gamma_{c1} = 2/M$---a straightforward calculation shows that it must be within $o(1/M^{5/2})$.

	For the second stage of the algorithm, we take the leading-order Hamiltonian $H^{(0)}$ to only include edges of weight $\Theta(M)$, and its diagram is shown in Figure~\ref{fig:simplex_two2_diagrams}c. When $\gamma$ takes its critical value of
	\[ \gamma_{c2} = \frac{1}{M}, \]
	the eigenstates $\ket{a}$, $\ket{b}$, and $\ket{g}$ of $H^{(0)}$ are triply degenerate. Then the perturbation $H^{(1)}$, which restores terms $\Theta(\sqrt{M})$, causes certain linear combinations $\alpha_a \ket{a} + \alpha_b \ket{b} + \alpha_g \ket{g}$ of them to be eigenstates of $H^{(0)} + H^{(1)}$. Doing the perturbative calculation to find the coefficients (as in the first case), the perturbed eigenstates for large $N$ are
	\begin{gather*}
		\frac{1}{\sqrt{2}} ( \ket{b} + \ket{a} ), \quad E = -1 - \frac{1}{\sqrt{M}} \\
		\ket{g}, \quad E = -1 \\
		\frac{1}{\sqrt{2}} ( \ket{b} - \ket{a} ), \quad E = -1 + \frac{1}{\sqrt{M}}.
	\end{gather*}
	So the system evolves from $\ket{b}$ to $\ket{a}$ in time $\pi / \mathrm{\Delta} E$:
	\[ t_2 = \frac{\pi \sqrt{M}}{2}. \]
	Diagrammatically, the perturbation $H^{(1)}$ restores edges $\Theta(\sqrt{M})$ in Figure~\ref{fig:simplex_two2_diagrams}c, and probability flows between $\ket{b}$ and $\ket{a}$.

	We can again use the method of \cite{Wong2015e} to find how precisely $\gamma$ must be chosen to its critical value $\gamma_{c2} = 1/M$---a straightforward calculation shows that it must be within $o(1/M^{3/2})$.

	These $\gamma_c$'s and runtimes are in agreement with Table~\ref{table:simplex_two}. 

\subsection{Two Marked, Case 3}

	As shown in Figure~\ref{fig:simplex_two3}, the system evolves in an 8-dimensional subspace spanned by
	\begin{align*}
		&\ket{a} = \frac{1}{\sqrt{2}} \sum_{i \in \text{red}} \ket{i}, &&\ket{e} = \frac{1}{\sqrt{2(M-2)}} \sum_{i \in \text{green}} \ket{i},\\
		&\ket{b} = \frac{1}{\sqrt{2(M-2)}} \sum_{i \in \text{blue}} \ket{i}, &&\ket{f} = \frac{1}{\sqrt{M-2}} \sum_{i \in \text{brown}} \ket{i}\\
		&\ket{c} = \frac{1}{\sqrt{2}} \sum_{i \in \text{yellow}} \ket{i}, &&\ket{g} = \frac{1}{\sqrt{(M-2)(M-3)}} \sum_{i \in \text{white}} \ket{i}, \\
		&\ket{d} = \frac{1}{\sqrt{M-2}} \sum_{i \in \text{magenta}} \ket{i}, &&\ket{i} = \frac{1}{\sqrt{2}} \sum_{i \in \text{gray}} \ket{i}.
	\end{align*}
	Note there are no $h$ type vertices. Instead, there's a new type, which we call $i$. In this subspace, the search Hamiltonian \eqref{eq:H} is
	\[ H = -\gamma \begin{pmatrix}
		\frac{1}{\gamma} & \sqrt{M_2} & 1 & 0 & 0 & 0 & 0 & 1 \\
		\sqrt{M_2} & M_3 & 0 & 0 & 1 & 0 & 0 & \sqrt{M_2} \\
		1 & 0 & 1 & \sqrt{2M_2} & 0 & 0 & 0 & 0 \\
		0 & 0 & \sqrt{2M_2} & M_3 & 0 & 1 & 0 & 0 \\
		0 & 1 & 0 & 0 & 1 & \sqrt{2} & \sqrt{2M_3} & 0 \\
		0 & 0 & 0 & 1 & \sqrt{2} & 0 & \sqrt{M_3} & 0 \\
		0 & 0 & 0 & 0 & \sqrt{2M_3} & \sqrt{M_3} & M_3 & 0 \\
		1 & \sqrt{M_2} & 0 & 0 & 0 & 0 & 0 & 1 \\
	\end{pmatrix}, \]
	where $M_2 = M-2$ and $M_3 = M-3$.

	\begin{figure}
	\begin{center}
		\subfloat[] {
			\includegraphics[scale=0.7]{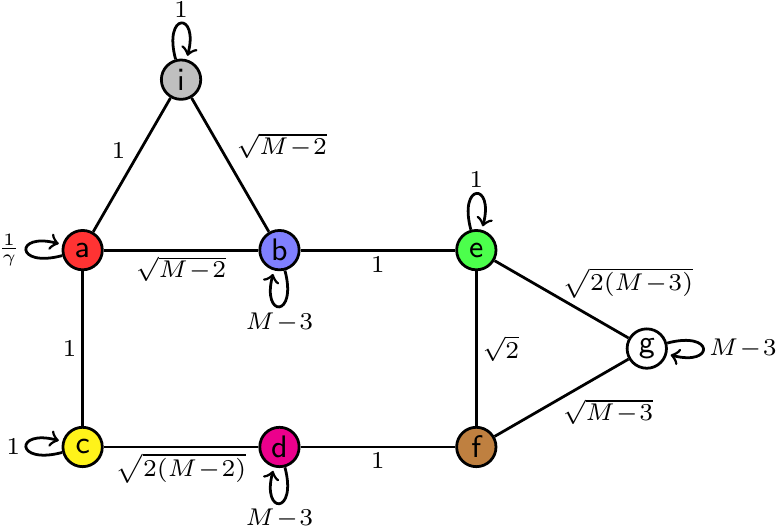}
		}
		\subfloat[] {
			\includegraphics[scale=0.7]{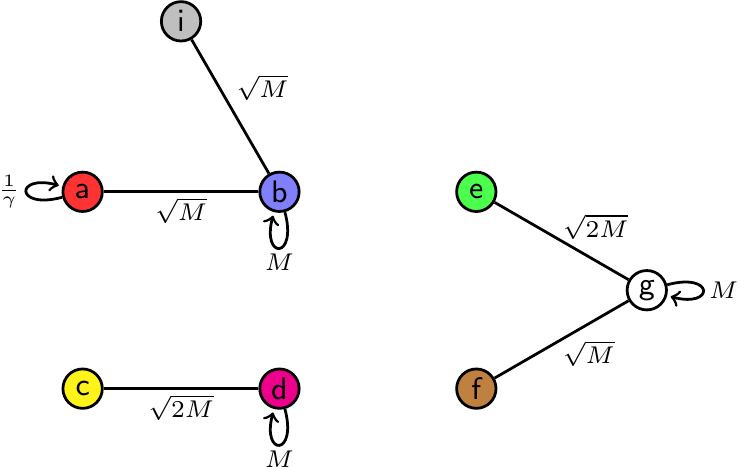}
		}
		\caption{\label{fig:simplex_two3_diagrams}Apart from a factor of $-\gamma$, \textbf{(a)} the Hamiltonian for the third case of search on the simplex of complete graphs with $k = 2$ marked vertices, and \textbf{(b)} the leading-order terms for the first stage of the algorithm.}
	\end{center}
	\end{figure}

	This Hamiltonian can be visualized as shown in Figure~\ref{fig:simplex_two3_diagrams}a. For the first stage of the algorithm, the leading-order Hamiltonian $H^{(0)}$ excludes edges that scale less than $\sqrt{M}$, and it can be visualized as shown in Figure~\ref{fig:simplex_two3_diagrams}b. As with the last two cases, there are two eigenvectors that we want to be degenerate. The first is
	\[ u = \frac{2 \sqrt{2}}{\sqrt{M}+\sqrt{M+12}} \ket{e} + \frac{2}{\sqrt{M}+\sqrt{M+12}} \ket{f} + \ket{g} \]
	with corresponding eigenvalue
	\[ E_u = \frac{-\gamma}{2} \left(M + \sqrt{M(M+12)} \right). \]
	The second eigenvector is messy, but can be approximated nicely. The leading-order Hamiltonian corresponding to $\ket{a}$, $\ket{b}$, and $\ket{i}$ is
	\[ H_{a,b,i}^{(0)} = -\gamma \begin{pmatrix}
		\frac{1}{\gamma} & \sqrt{M} & 0 \\
		\sqrt{M} & M & \sqrt{M} \\
		0 & \sqrt{M} & 0 \\
	\end{pmatrix}. \]
	The eigenvalues $\lambda$ of this satisfy the characteristic equation
	\[ -\lambda^3 - (\gamma M + 1)\lambda^2 + \gamma M(2\gamma - 1)\lambda + \gamma^2 M = 0. \]
	When $\gamma$ takes its critical value of
	\[ \gamma_{c1} = \frac{2}{M} - \frac{6}{M^2} + \frac{36}{M^3}, \]
	one of these eigenvalues, which we will call $E_v$, and $E_u$ both equal $-2 - 270/M^3 + O(1/M^4)$, making them approximately degenerate. To find the corresponding eigenvector $v$, we use the first and third lines of the eigenvalue equation $H_{a,b,i}^{(0)} v = E_v v$:
	\[ -\gamma \begin{pmatrix}
		\frac{1}{\gamma} & \sqrt{M} & 0 \\
		\sqrt{M} & M & \sqrt{M} \\
		0 & \sqrt{M} & 0 \\
	\end{pmatrix} \begin{pmatrix}
		v_a \\ v_b \\ v_i
	\end{pmatrix} = E_v \begin{pmatrix}
		v_a \\ v_b \\ v_i
	\end{pmatrix}. \]
	This yields
	\[ v = \frac{-\gamma \sqrt{M}}{E_v + 1} \ket{a} + \ket{b} + \frac{-\gamma \sqrt{M}}{E_v} \ket{i}. \]
	The perturbation $H^{(1)}$, which restores terms of constant weight, causes certain linear combinations $\alpha_u \ket{u} + \alpha_v \ket{v}$ to be eigenstates of $H^{(0)} + H^{(1)}$. Doing the perturbative calculation to find the coefficients, the perturbed eigenstates for large $N$ are
	\begin{gather*}
		\frac{1}{\sqrt{2}} ( \ket{u} + \ket{v} ), \quad E = -2 - \frac{2\sqrt{2}}{M^{3/2}} \\
		\frac{1}{\sqrt{2}} ( \ket{u} - \ket{v} ), \quad E = -2 + \frac{2\sqrt{2}}{M^{3/2}}.
	\end{gather*}
	Since $\ket{u} \approx \ket{g}$ and $\ket{v} \approx \ket{b}$ for large $N$, the system evolves from $\ket{s} \approx \ket{g}$ to $\ket{b}$ in time $\pi / \mathrm{\Delta} E$:
	\[ t_1 = \frac{\pi M^{3/2}}{4\sqrt{2}}. \]
	Diagrammatically, the perturbation $H^{(1)}$ restores edges of constant weight in Figure~\ref{fig:simplex_two3_diagrams}b, and probability flows between $\ket{g}$ and $\ket{b}$ since they are the most dominant terms.

	We can again use the method of \cite{Wong2015e} to find how precisely $\gamma$ must be chosen to its critical value $\gamma_{c1} = 2/M$---a straightforward calculation shows that it must be within $o(1/M^{5/2})$.

	The second stage of the algorithm is similar to the previous two cases, where we take the leading-order Hamiltonian $H^{(0)}$ to only include edges of weight $\Theta(M)$. When $\gamma$ takes its critical value of
	\[ \gamma_{c2} = \frac{1}{M}, \]
	then $\ket{a}$ and $\ket{b}$ are degenerate eigenvectors (among others) of $H^{(0)}$. The perturbation $H^{(1)}$ restores terms of order $\Theta(\sqrt{M})$, which causes probability to flow between $\ket{b}$ and $\ket{a}$. Doing the calculation, we find eigenstates of the perturbed system that are proportional to $\ket{b} \pm \ket{a}$ with eigenvalues $-1 \mp 1 / \sqrt{M}$, so the system evolves from $\ket{b}$ to $\ket{a}$ in time $\pi / \mathrm{\Delta} E$:
	\[ t_2 = \frac{\pi \sqrt{M}}{2}. \]

	As before, a straightforward calculation using the method of \cite{Wong2015e} shows that $\gamma$ must be chosen within $o(1/M^{3/2})$ of its critical value $\gamma_{c2} = 1/M$.

	These $\gamma_c$'s and runtimes are in agreement with Table~\ref{table:simplex_two}. 

\subsection{Two Marked, Case 4}

	As shown in Figure~\ref{fig:simplex_two4}, the system evolves in an 11-dimensional subspace spanned by
	\begin{align*}
		&\ket{a} = \frac{1}{\sqrt{2}} \sum_{i \in \text{red}} \ket{i}, &&\ket{g} = \frac{1}{\sqrt{(M-3)(M-4)}} \sum_{i \in \text{white}} \ket{i}, \\
		&\ket{b} = \frac{1}{\sqrt{2(M-3)}} \sum_{i \in \text{blue}} \ket{i}, &&\ket{h} = \frac{1}{\sqrt{2}} \sum_{i \in \text{orange}} \ket{i}, \\
		&\ket{c} = \frac{1}{\sqrt{2}} \sum_{i \in \text{yiellow}} \ket{i}, &&\ket{i} = \frac{1}{\sqrt{2}} \sum_{i \in \text{gray}} \ket{i}, \\
		&\ket{d} = \frac{1}{\sqrt{2(M-3)}} \sum_{i \in \text{magenta}} \ket{i}, &&\ket{j} = \frac{1}{\sqrt{2}} \sum_{i \in \text{pink}} \ket{i}, \\
		&\ket{e} = \frac{1}{\sqrt{2(M-3)}} \sum_{i \in \text{green}} \ket{i}, &&\ket{k} = \frac{1}{\sqrt{2}} \sum_{i \in \text{lime}} \ket{i}. \\
		&\ket{f} = \frac{1}{\sqrt{2(M-3)}} \sum_{i \in \text{brown}} \ket{i},
	\end{align*}
	In this subspace, the search Hamiltonian \eqref{eq:H} is
	\setcounter{MaxMatrixCols}{11}
	\[ H = -\gamma \begin{pmatrix}
		\frac{1}{\gamma} & \sqrt{M_3} & 1 & 0 & 0 & 0 & 0 & 0 & 1 & 1 & 0 \\
		\sqrt{M_3} & M_4 & 0 & 0 & 1 & 0 & 0 & 0 & \sqrt{M_3} & \sqrt{M_3} & 0 \\
		1 & 0 & 0 & \sqrt{M_3} & 0 & 0 & 0 & 1 & 0 & 0 & 1 \\
		0 & 0 & \sqrt{M_3} & M_4 & 0 & 1 & 0 & \sqrt{M_3} & 0 & 0 & \sqrt{M_3} \\
		0 & 1 & 0 & 0 & 1 & 2 & \sqrt{2M_4} & 0 & 0 & 0 & 0 \\
		0 & 0 & 0 & 1 & 2 & 1 & \sqrt{2M_4} & 0 & 0 & 0 & 0 \\
		0 & 0 & 0 & 0 & \sqrt{2M_4} & \sqrt{2M_4} & M_4 & 0 & 0 & 0 & 0 \\
		0 & 0 & 1 & \sqrt{M_3} & 0 & 0 & 0 & 1 & 0 & 0 & 1 \\
		1 & \sqrt{M_3} & 0 & 0 & 0 & 0 & 0 & 0 & 1 & 1 & 0 \\
		1 & \sqrt{M_3} & 0 & 0 & 0 & 0 & 0 & 0 & 1 & 0 & 1 \\
		0 & 0 & 1 & \sqrt{M_3} & 0 & 0 & 0 & 1 & 0 & 1 & 0 \\
	\end{pmatrix}, \]
	where $M_3 = M-3$ and $M_4 = M-4$.

	\begin{figure}
	\begin{center}
		\subfloat[] {
			\includegraphics[scale=0.7]{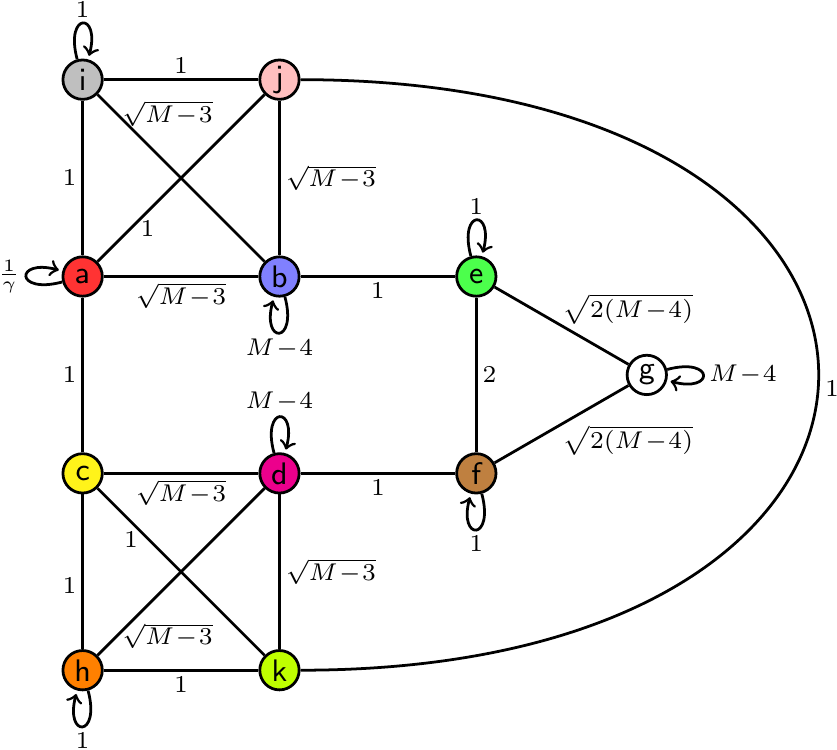}
		}
		\subfloat[] {
			\includegraphics[scale=0.7]{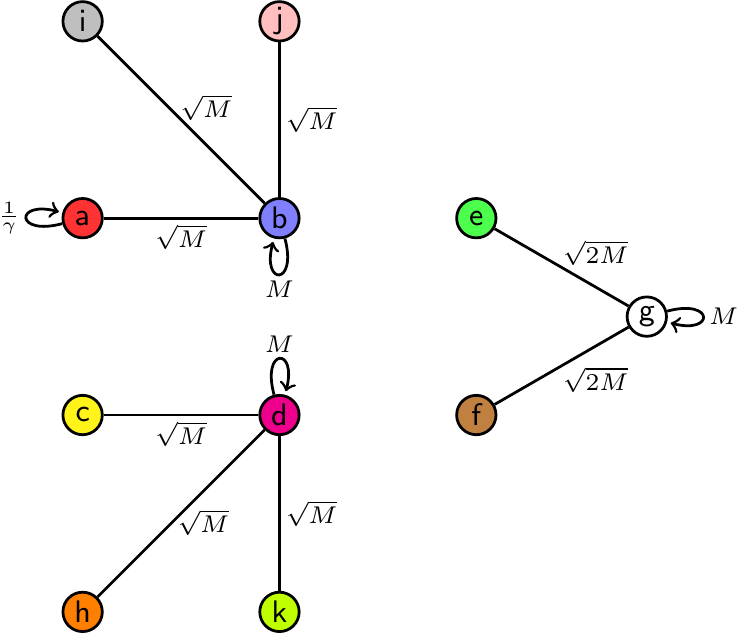}
		}
		\caption{\label{fig:simplex_two4_diagrams}Apart from a factor of $-\gamma$, \textbf{(a)} the Hamiltonian for the fourth case of search on the simplex of complete graphs with $k = 2$ marked vertices, and \textbf{(b)} the leading-order terms for the first stage of the algorithm.}
	\end{center}
	\end{figure}

	This Hamiltonian can be visualized as shown in Figure~\ref{fig:simplex_two4_diagrams}a. For the first stage of the algorithm, the leading-order Hamiltonian $H^{(0)}$ excludes edges that scale less than $\sqrt{M}$, and it can be visualized as shown in Figure~\ref{fig:simplex_two4_diagrams}b. As with the last two cases, there are two eigenvectors that we want to be degenerate. The first is
	\[ u = \frac{2\sqrt{2}}{\sqrt{M} + \sqrt{M+16}} \ket{e} + \frac{2\sqrt{2}}{\sqrt{M} + \sqrt{M+16}} \ket{f} + \ket{g}, \]
	with corresponding eigenvalue
	\[ E_u = \frac{-\gamma}{2} \left( M + \sqrt{M}\sqrt{M+16} \right). \]
	The second eigenvector is messy, but can be approximated nicely. The leading-order Hamiltonian corresponding to $\ket{a}$, $\ket{b}$, $\ket{i}$, and $\ket{j}$ is
	\[ H_{a,b,i,j}^{(0)} = -\gamma \begin{pmatrix}
		\frac{1}{\gamma} & \sqrt{M} & 0 & 0 \\
		\sqrt{M} & M & \sqrt{M} & \sqrt{M} \\
		0 & \sqrt{M} & 0 & 0 \\
		0 & \sqrt{M} & 0 & 0 \\
	\end{pmatrix}. \]
	The eigenvalues $\lambda$ of this satisfy the characteristic equation
	\[ \lambda^4 + (1 + M\gamma)\lambda^3 + M\gamma(1-3\gamma)\lambda^2 - 2M\gamma^2 \lambda = 0. \]
	When $\gamma$ takes its critical value of
	\[ \gamma_{c1} = \frac{2}{M} - \frac{8}{M^2} + \frac{64}{M^3}, \]
	one of these eigenvalues, which we will call $E_v$, and $E_u$ both equal $-2 - 640/M^3 - O(1/M^4)$, making them approximately degenerate. To find the corresponding eigenvector $v$, we use the eigenvalue equation $H_{a,b,i,j}^{(0)} v = E_v v$:
	\[ -\gamma \begin{pmatrix}
		\frac{1}{\gamma} & \sqrt{M} & 0 & 0 \\
		\sqrt{M} & M & \sqrt{M} & \sqrt{M} \\
		0 & \sqrt{M} & 0 & 0 \\
		0 & \sqrt{M} & 0 & 0 \\
	\end{pmatrix} \begin{pmatrix}
		v_a \\ v_b \\ v_i \\ v_j
	\end{pmatrix} = E_v \begin{pmatrix}
		v_a \\ v_b \\ v_i \\ v_j
	\end{pmatrix}. \]
	This yields
	\[ v = \frac{-\gamma \sqrt{M}}{1+E} \ket{a} + \ket{b} - \frac{\gamma \sqrt{M}}{E} \ket{i} - \frac{\gamma \sqrt{M}}{E} \ket{j}. \]
	The perturbation $H^{(1)}$, which restores terms of constant weight, causes certain linear combinations $\alpha_u \ket{u} + \alpha_v \ket{v}$ to be eigenstates of $H^{(0)} + H^{(1)}$. Doing the perturbative calculation to find the coefficients, the perturbed eigenstates for large $N$ are
	\begin{gather*}
		\frac{1}{\sqrt{2}} ( \ket{u} + \ket{v} ), \quad E = -2 - \frac{2\sqrt{2}}{M^{3/2}} \\
		\frac{1}{\sqrt{2}} ( \ket{u} - \ket{v} ), \quad E = -2 + \frac{2\sqrt{2}}{M^{3/2}}.
	\end{gather*}
	Since $\ket{u} \approx \ket{g}$ and $\ket{v} \approx \ket{b}$ for large $N$, the system evolves from $\ket{s} \approx \ket{g}$ to $\ket{b}$ in time $\pi / \mathrm{\Delta} E$:
	\[ t_1 = \frac{\pi M^{3/2}}{4\sqrt{2}}. \]
	Diagrammatically, the perturbation $H^{(1)}$ restores edges of constant weight in Figure~\ref{fig:simplex_two4_diagrams}b, and probability flows between $\ket{g}$ and $\ket{b}$ since they are the most dominant terms.

	We can again use the method of \cite{Wong2015e} to find how precisely $\gamma$ must be chosen to its critical value $\gamma_{c1} = 2/M$---a straightforward calculation shows that it must be within $o(1/M^{5/2})$.

	The second stage of the algorithm is similar to the previous three cases, where we take the leading-order Hamiltonian $H^{(0)}$ to only include edges of weight $\Theta(M)$. When $\gamma$ takes its critical value of
	\[ \gamma_{c2} = \frac{1}{M}, \]
	then $\ket{a}$ and $\ket{b}$ are degenerate eigenvectors (among others) of $H^{(0)}$. The perturbation $H^{(1)}$ restores terms of order $\Theta(\sqrt{M})$, which causes probability to flow between $\ket{b}$ and $\ket{a}$. Doing the calculation, we find eigenstates of the perturbed system that are proportional to $\ket{b} \pm \ket{a}$ with eigenvalues $-1 \mp 1 / \sqrt{M}$, so the system evolves from $\ket{b}$ to $\ket{a}$ in time $\pi / \mathrm{\Delta} E$:
	\[ t_2 = \frac{\pi \sqrt{M}}{2}. \]

	As before, a straightforward calculation using the method of \cite{Wong2015e} shows that $\gamma$ must be chosen within $o(1/M^{3/2})$ of its critical value $\gamma_{c2} = 1/M$.

	These $\gamma_c$'s and runtimes are in agreement with Table~\ref{table:simplex_two}. 

\subsection{Two Marked, Case 5}

	As shown in Figure~\ref{fig:simplex_two5}, the system evolves in a 13-dimensional subspace spanned by
	\begin{align*}
		&\ket{a} = \ket{\text{red}}, && \ket{h} = \frac{1}{\sqrt{M-2}} \sum_{i \in \text{orange}} \ket{i}, \\
		&\ket{b} = \frac{1}{\sqrt{M-2}} \sum_{i \in \text{blue}} \ket{i}, && \ket{i} = \ket{\text{gray}}, \\
		&\ket{c} = \ket{\text{yellow}}, && \ket{j} = \ket{\text{pink}}, \\
		&\ket{d} = \ket{\text{magenta}}, && \ket{k} = \frac{1}{\sqrt{M-2}} \sum_{i \in \text{lime}} \ket{i}, \\
		&\ket{e} = \frac{1}{\sqrt{M-2}} \sum_{i \in \text{green}} \ket{i}, && \ket{l} = \frac{1}{\sqrt{M-2}} \sum_{i \in \text{teal}} \ket{i}, \\
		&\ket{f} = \ket{\text{brown}}, && \ket{m} = \frac{1}{\sqrt{(M-2)(M-3)}} \sum_{i \in \text{violet}} \ket{i}. \\
		&\ket{g} = \frac{1}{\sqrt{M-2}} \sum_{i \in \text{white}} \ket{i},
	\end{align*}
	In this subspace, the search Hamiltonian \eqref{eq:H} is
	\setcounter{MaxMatrixCols}{13}
	\[ \setlength{\arraycolsep}{1pt} H = -\gamma \begin{pmatrix}
		\frac{1}{\gamma} & \sqrt{M_2} & 1 & 0 & 0 & 0 & 0 & 0 & 1 & 0 & 0 & 0 & 0 \\
		\sqrt{M_2} & M_3 & 0 & 0 & 1 & 0 & 0 & 0 & \sqrt{M_2} & 0 & 0 & 0 & 0 \\
		1 & 0 & 0 & 1 & 0 & 0 & 0 & \sqrt{M_2} & 0 & 0 & 0 & 0 & 0 \\
		0 & 0 & 1 & \frac{1}{\gamma} & 0 & 1 & 0 & \sqrt{M_2} & 0 & 0 & 0 & 0 & 0 \\
		0 & 1 & 0 & 0 & 0 & 0 & 0 & 0 & 0 & 0 & 1 & 1 & \sqrt{M_3} \\
		0 & 0 & 0 & 1 & 0 & 0 & \sqrt{M_2} & 0 & 0 & 1 & 0 & 0 & 0 \\
		0 & 0 & 0 & 0 & 0 & \sqrt{M_2} & M_3 & 0 & 0 & \sqrt{M_2} & 1 & 0 & 0 \\
		0 & 0 & \sqrt{M_2} & \sqrt{M_2} & 0 & 0 & 0 & M_3 & 0 & 0 & 0 & 1 & 0 \\
		1 & \sqrt{M_2} & 0 & 0 & 0 & 0 & 0 & 0 & 0 & 1 & 0 & 0 & 0 \\
		0 & 0 & 0 & 0 & 0 & 1 & \sqrt{M_2} & 0 & 1 & 0 & 0 & 0 & 0 \\
		0 & 0 & 0 & 0 & 1 & 0 & 1 & 0 & 0 & 0 & 0 & 1 & \sqrt{M_3} \\
		0 & 0 & 0 & 0 & 1 & 0 & 0 & 1 & 0 & 0 & 1 & 0 & \sqrt{M_3} \\
		0 & 0 & 0 & 0 & \sqrt{M_3} & 0 & 0 & 0 & 0 & 0 & \sqrt{M_3} & \sqrt{M_3} & M_3 \\
	\end{pmatrix}, \]
	where $M_2 = M-2$ and $M_3 = M-3$.
	
	\begin{figure}
	\begin{center}
		\subfloat[] {
			\includegraphics[scale=0.65]{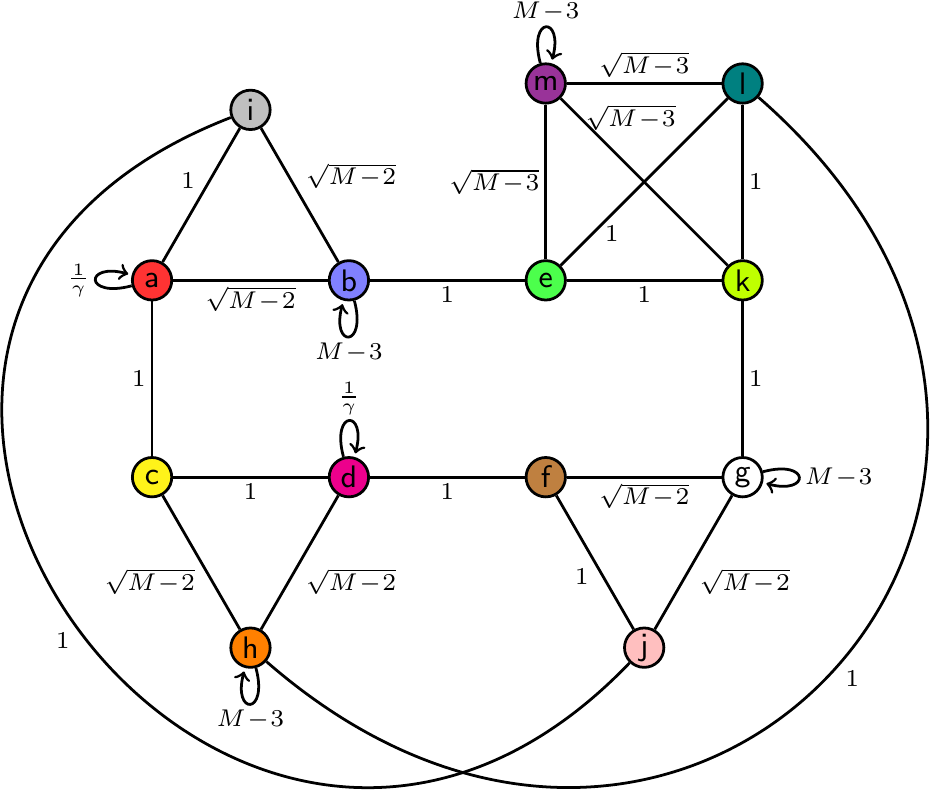}
		}
		\subfloat[] {
			\includegraphics[scale=0.65]{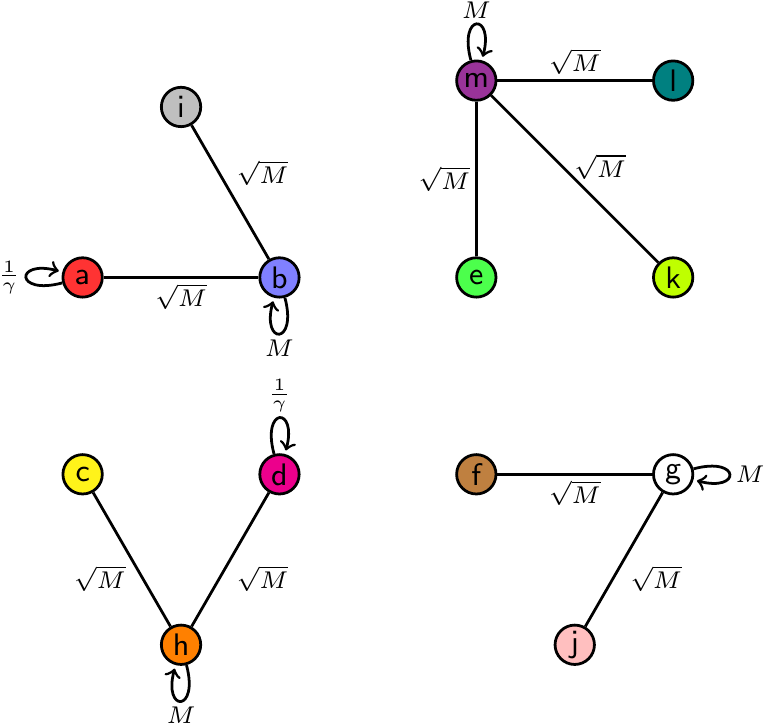}
		}
		\caption{\label{fig:simplex_two5_diagrams}Apart from a factor of $-\gamma$, \textbf{(a)} the Hamiltonian for the fifth case of search on the simplex of complete graphs with $k = 2$ marked vertices, and \textbf{(b)} the leading-order terms for the first stage of the algorithm.}
	\end{center}
	\end{figure}

	This Hamiltonian can be visualized as shown in Figure~\ref{fig:simplex_two5_diagrams}a. For the first stage of the algorithm, the leading-order Hamiltonian $H^{(0)}$ excludes edges that scale less than $\sqrt{M}$, and it can be visualized as shown in Figure~\ref{fig:simplex_two5_diagrams}b. The initial equal superposition state $\ket{s}$ is approximately $\ket{m}$ for large $N$, and we want it to evolve to the marked vertices $\ket{a}$ and $\ket{d}$. So we will need leading-order eigenstates that are approximately each of these to be triply degenerate. The first is
	\[ u = \frac{2}{\sqrt{M}+\sqrt{12+M}} ( \ket{e} + \ket{k} + \ket{l} ) + \ket{m} \]
	with corresponding eigenvalue
	\[ E_u = \frac{-\gamma}{2} \left(M + \sqrt{M(M+12)} \right). \]
	Note this is the same eigenvalue as $E_u$ from Case 3. For the other two leading-order eigenstates, Figure~\ref{fig:simplex_two5_diagrams}b reveals that $H_{a,b,i}^{(0)}$ and $H_{c,d,h}^{(0)}$ are identical with $a \sim d$, $b \sim h$, and $i \sim c$, so their corresponding eigenstates are always degenerate. Furthermore, they are identical to $H_{a,b,i}^{(0)}$ from Figure~\ref{fig:simplex_two3_diagrams}b from Case 3. So the eigenvectors and eigenvalues carry over:
	\begin{gather*}
		v = \frac{-\gamma \sqrt{M}}{E_v + 1} \ket{a} + \ket{b} + \frac{-\gamma \sqrt{M}}{E_v} \ket{i}, \\
		w = \frac{-\gamma \sqrt{M}}{E_w + 1} \ket{d} + \ket{h} + \frac{-\gamma \sqrt{M}}{E_w} \ket{c},
	\end{gather*}
	including the critical $\gamma$
	\[ \gamma_{c1} = \frac{2}{M} - \frac{6}{M^2} + \frac{36}{M^3}, \]
	at which the eigenvalues $E_u$, $E_v$, and $E_w$ all equal $-2 - 270/M^3 + O(1/M^4)$, making them approximately degenerate.
	
	With the perturbation $H^{(1)}$, which restores terms of constant weight, we have the same behavior and runtime
	\[ t_1 = \frac{\pi M^{3/2}}{4\sqrt{2}} \]
	as Case 3, except the system evolves from $\ket{s} \approx \ket{m}$ to $\ket{b} + \ket{h}$. So the probability gets split between the two paths. Diagrammatically, the perturbation $H^{(1)}$ restores edges of constant weight in Figure~\ref{fig:simplex_two5_diagrams}b, and probability flows from $\ket{m}$ to $\ket{b}$ and $\ket{h}$ since they are the most dominant terms.

	We can again use the method of \cite{Wong2015e} to find how precisely $\gamma$ must be chosen to its critical value $\gamma_{c2} = 1/M$---a straightforward calculation shows that it must be within $o(1/M^{5/2})$.

	The second stage of the algorithm is similar to the previous cases, where we take the leading-order Hamiltonian $H^{(0)}$ to only include edges of weight $\Theta(M)$. When $\gamma$ takes its critical value of
	\[ \gamma_{c2} = \frac{1}{M}, \]
	then $\ket{a}$ and $\ket{b}$ are degenerate eigenvectors, as are $\ket{d}$ and $\ket{h}$, of $H^{(0)}$. The perturbation $H^{(1)}$ restores terms of order $\Theta(\sqrt{M})$, which causes probability to flow from $\ket{b}$ to $\ket{a}$ and from $\ket{h}$ to $\ket{d}$. The runtime from Case 3 carries over:
	\[ t_2 = \frac{\pi \sqrt{M}}{2}. \]

	As before, a straightforward calculation using the method of \cite{Wong2015e} shows that $\gamma$ must be chosen within $o(1/M^{3/2})$ of its critical value $\gamma_{c2} = 1/M$.

	These $\gamma_c$'s and runtimes are in agreement with Table~\ref{table:simplex_two}. 

\section{\label{appendix:larger}Details for Larger Examples}

In this appendix, we employ degenerate perturbation theory \cite{Sakurai1993,JMW2014,Wong2014} to find the critical $\gamma$'s and runtimes for search with a larger number of marked vertices, the results which are summarized in Table~\ref{table:simplex_larger}.

\subsection{One Marked Per Complete}

	As shown in Figure~\ref{fig:simplex_M1}a, the system evolves in a 3-dimensional subspace spanned by
	\begin{gather*}
		\ket{a} = \frac{1}{\sqrt{M+1}} \sum_{i \in \text{red}} \ket{i}, \\
		\ket{b} = \frac{1}{\sqrt{(M+1)(M-2)}} \sum_{i \in \text{blue}} \ket{i}, \\
		\ket{c} = \frac{1}{\sqrt{M+1}} \sum_{i \in \text{yellow}} \ket{i}.
	\end{gather*}
	In this subspace, the search Hamiltonian \eqref{eq:H} is
	\[ H = -\gamma \begin{pmatrix}
		\frac{1}{\gamma} & \sqrt{M-2} & 2 \\
		\sqrt{M-2} & M-2 & \sqrt{M-2} \\
		2 & \sqrt{M-2} & 0 \\
	\end{pmatrix}. \]

	\begin{figure}
	\begin{center}
		\subfloat[] {
			\includegraphics[scale=0.7]{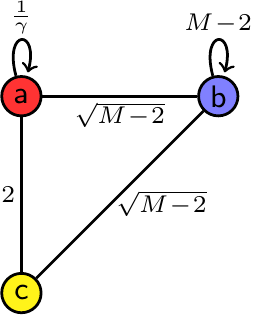}
		} \quad \quad \quad \quad \quad \quad 
		\subfloat[] {
			\includegraphics[scale=0.7]{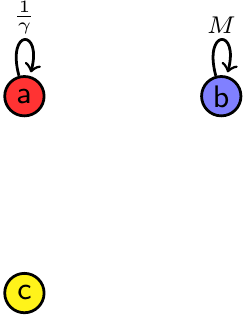}
		}
		\caption{\label{fig:simplex_M1a_diagrams}Apart from a factor of $-\gamma$, \textbf{(a)} the Hamiltonian for the first case of search on the simplex of complete graphs with $k = M+1$ marked vertices, and \textbf{(b)} the leading-order terms. }
	\end{center}
	\end{figure}

	This Hamiltonian can be visualized as shown in Figure~\ref{fig:simplex_M1a_diagrams}a. The leading-order Hamiltonian $H^{(0)}$ excludes edges that scale less than $M$, and it can be visualized as shown in Figure~\ref{fig:simplex_M1a_diagrams}b. Clearly, the eigenstates of this are $\ket{a}$, $\ket{b}$, and $\ket{c}$ with corresponding eigenvalues $-1$, $-\gamma M$, and $0$. Since $\ket{s} \approx \ket{b}$, we choose $\gamma$ so that $\ket{b}$ and $\ket{a}$ are degenerate, \textit{i.e.},
	\[ \gamma_c = \frac{1}{M}. \]
	The perturbation $H^{(1)}$, which restores edges of weight $\Theta(\sqrt{M})$, causes certain linear combinations $\alpha_a \ket{a} + \alpha_b \ket{b}$ to be eigenstates of $H^{(0)} + H^{(1)}$. The coefficients can be found by solving
	\[ \begin{pmatrix}
		H_{aa} & H_{ab} \\
		H_{ba} & H_{bb} \\
	\end{pmatrix} \begin{pmatrix}
		\alpha_a \\
		\alpha_b \\
	\end{pmatrix} = E \begin{pmatrix}
		\alpha_a \\
		\alpha_b \\
	\end{pmatrix}, \]
	where $H_{ab} = \langle a | H^{(0)} + H^{(1)} | b \rangle$, \textit{etc}. With $\gamma = \gamma_c$, this yields eigenstates and eigenvalues
	\begin{gather*}
		\ket{\psi_0} = \frac{1}{\sqrt{2}} ( \ket{b} + \ket{a} ), \quad E_0 = -1 - \frac{1}{\sqrt{M}} \\
		\ket{\psi_1} = \frac{1}{\sqrt{2}} ( \ket{b} - \ket{a} ), \quad E_1 = - 1 + \frac{1}{\sqrt{M}}.
	\end{gather*}
	So the system evolves from $\ket{s} \approx \ket{b}$ to $\ket{a}$ in time $\pi/\mathrm{\Delta} E$:
	\[ t_* = \frac{\pi \sqrt{M}}{2}. \]

	Using the method Section VI of \cite{Wong2015e}, if $\gamma$ is within $\epsilon$ of its critical value of $\gamma_c = 1/M$, then the eigenvalue of $\ket{b}$ is now $-\gamma M = -1 - \epsilon M$. In the perturbative calculation, this introduces a leading-order (in $\epsilon$) term $\epsilon M$ due to $H_{bb}$. For this to not influence the energy gap $\Theta(1/\sqrt{M})$, we require $\epsilon M = o(1/\sqrt{M})$, or $\epsilon = o(1/M^{3/2})$. Thus for the algorithm to asymptotically evolve from $\ket{b}$ to $\ket{a}$, we require $\gamma = \gamma_c + o(1/M^{3/2})$.

	This $\gamma_c$ and runtime are in agreement with Table~\ref{table:simplex_larger}. 
	
\subsection{Fully Marked Complete, Plus One}

	As shown in Figure~\ref{fig:simplex_M1}b, the system evolves in a 7-dimensional subspace spanned by
	\begin{align*}
		&\ket{a} = \ket{\text{red}}, &&\ket{e} = \frac{1}{\sqrt{M-1}} \sum_{i \in \text{green}} \ket{i}, \\
		&\ket{b} = \frac{1}{\sqrt{M-1}} \sum_{i \in \text{blue}} \ket{i}, &&\ket{f} = \frac{1}{\sqrt{M-1}} \sum_{i \in \text{brown}} \ket{i}, \\
		&\ket{c} = \ket{\text{yellow}}, &&\ket{g} = \frac{1}{\sqrt{(M-1)(M-2)}} \sum_{i \in \text{white}} \ket{i}. \\
		&\ket{d} = \frac{1}{\sqrt{M-1}} \sum_{i \in \text{magenta}} \ket{i},
	\end{align*}
	In this subspace, the search Hamiltonian \eqref{eq:H} is
	\[ H = -\gamma \begin{pmatrix}
		\frac{1}{\gamma} & \sqrt{M-1} & 1 & 0 & 0 & 0 & 0 \\
		\sqrt{M-1} & M-2+\frac{1}{\gamma} & 0 & 0 & 1 & 0 & 0 \\
		1 & 0 & \frac{1}{\gamma} & \sqrt{M-1} & 0 & 0 & 0 \\
		0 & 0 & \sqrt{M-1} & M-2 & 0 & 1 & 0 \\
		0 & 1 & 0 & 0 & 0 & 1 & \sqrt{M-2} \\
		0 & 0 & 0 & 1 & 1 & 0 & \sqrt{M-2} \\
		0 & 0 & 0 & 0 & \sqrt{M-2} & \sqrt{M-2} & M-2 \\
	\end{pmatrix}. \]

	\begin{figure}
	\begin{center}
		\subfloat[] {
			\includegraphics[scale=0.7]{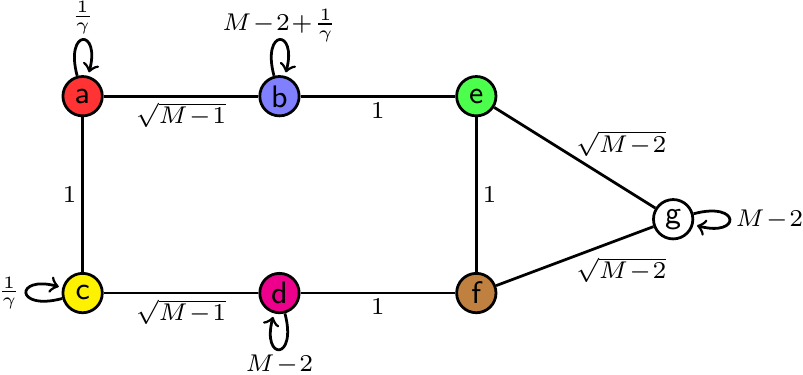}
		}
		\subfloat[] {
			\includegraphics[scale=0.7]{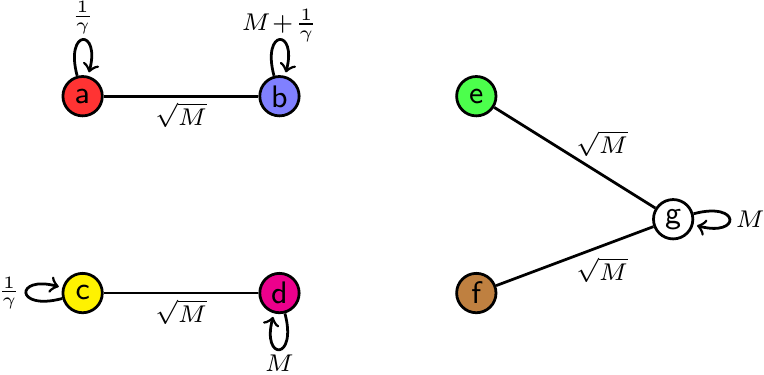}
		}
		\caption{\label{fig:simplex_M1b_diagrams}Apart from a factor of $-\gamma$, \textbf{(a)} the Hamiltonian for the second case of search on the simplex of complete graphs with $k = M+1$ marked vertices, and \textbf{(b)} the leading-order terms. }
	\end{center}
	\end{figure}

	This Hamiltonian can be visualized as shown in Figure~\ref{fig:simplex_M1b_diagrams}a. The leading-order Hamiltonian $H^{(0)}$ excludes edges that scale less than $\sqrt{M}$, and it can be visualized as shown in Figure~\ref{fig:simplex_M1b_diagrams}b. The two eigenstates of $H^{(0)}$ that we want to be degenerate are
	\[ u = \frac{2}{\sqrt{M}+\sqrt{8+M}} \ket{e} + \frac{2}{\sqrt{M}+\sqrt{8+M}} \ket{f} + \ket{g} \]
	with corresponding eigenvalue
	\[ E_u = \frac{-\gamma}{2} \left( M + \sqrt{M(M+8)} \right) \]
	and
	\[ v = \frac{1}{2} \left( \sqrt{M+4} - \sqrt{M} \right) \ket{a} + \ket{b} \]
	with corresponding eigenvalue
	\[ E_v = \frac{1}{2} \left(-2-M \gamma -\sqrt{M} \sqrt{M+4} \gamma \right). \]
	These are degenerate when $\gamma$ takes its critical value of
	\[ \gamma_c = \frac{2}{\sqrt{M} \left(\sqrt{M+8}-\sqrt{M+4}\right)} \approx 1 + \frac{3}{M} + O(1/M^2). \]
	The perturbation $H^{(1)}$, which restores terms of constant weight, causes certain linear combinations $\alpha_u \ket{u} + \alpha_v \ket{v}$ to be eigenstates of $H^{(0)} + H^{(1)}$. Doing the perturbative calculation to find the coefficients, the perturbed eigenstates for large $N$ are
	\[ \frac{1}{\sqrt{2}} \left( \ket{u} + \ket{v} \right), \quad E = -3 - \frac{1}{\sqrt{M}} \]
	\[ \frac{1}{\sqrt{2}} \left( \ket{u} - \ket{v} \right), \quad E = -3 + \frac{1}{\sqrt{M}} \]
	Since $\ket{u} \approx \ket{g}$ and $\ket{v}\approx \ket{b}$ for large $N$, the system evolves from $\ket{s} \approx \ket{g}$ to $\ket{b}$, which is marked, in time $\pi / \mathrm{\Delta} E$:
	\[ t_* = \frac{\pi \sqrt{M}}{2}. \]

	We can again use the method of \cite{Wong2015e} to find how precisely $\gamma$ must be chosen to its critical value $\gamma_{c} = 1 + 3/M$---a straightforward calculation shows that it must be within $o(1/M^{3/2})$.

	This $\gamma_c$ and runtime are in agreement with Table~\ref{table:simplex_larger}. 

\subsection{Two Marked Per Complete Graph, Case 1}

	As shown in Figure~\ref{fig:simplex_2M1}a, the system evolves in a 2-dimensional subspace spanned by
	\begin{gather*}
		\ket{a} = \frac{1}{\sqrt{2(M+1)}} \sum_{i \in \text{red}} \ket{i}, \\
		\ket{b} = \frac{1}{\sqrt{(M+1)(M-2)}} \sum_{i \in \text{blue}} \ket{i}.
	\end{gather*}
	In this subspace, the search Hamiltonian \eqref{eq:H} is
	\[ H = -\gamma \begin{pmatrix}
		2 + \frac{1}{\gamma} & \sqrt{2(M-2)} \\
		\sqrt{2(M-2)} & M-2 \\
	\end{pmatrix}. \]
	We can find the eigenvectors and eigenvalues of this directly without perturbation theory. They are
	\[ \psi_0 = \frac{1+4 \gamma -M \gamma -\sqrt{1+8 \gamma -2 M \gamma +M^2 \gamma ^2}}{2 \sqrt{2} \sqrt{M-2} \gamma } \ket{a} + \ket{b} \]
	with corresponding eigenvalue
	\[ E_0 = \frac{1}{2} \left(-1-M \gamma +\sqrt{1+8 \gamma -2 M \gamma +M^2 \gamma ^2}\right) \]
	and
	\[ \psi_1 = \frac{1+4 \gamma -M \gamma +\sqrt{1+8 \gamma -2 M \gamma +M^2 \gamma ^2}}{2 \sqrt{2} \sqrt{M-2} \gamma } \ket{a} + \ket{b} \]
	with corresponding eigenvalue
	\[ E_1 = \frac{1}{2} \left(-1-M \gamma -\sqrt{1+8 \gamma -2 M \gamma +M^2 \gamma ^2}\right). \]
	When $\gamma$ takes its critical value of
	\[ \gamma_c = \frac{1}{M}, \]
	these become for large $N$
	\[ \ket{\psi_0} = \frac{1}{\sqrt{2}} ( \ket{b} + \ket{a} ), \quad E_0 = -1 - \sqrt{\frac{2}{M}} \]
	\[ \ket{\psi_1} = \frac{1}{\sqrt{2}} ( \ket{b} - \ket{a} ), \quad E_1 = -1 + \sqrt{\frac{2}{M}} \]
	So the system evolves from $\ket{s} \approx \ket{b}$ to $\ket{a}$ in time $\pi / \mathrm{\Delta} E$:
	\[ t_* = \frac{\pi \sqrt{M}}{2 \sqrt{2}}. \]

	An explicit calculation as in \cite{Wong2015c} shows that $\gamma$ must be chosen within $o(1/M^{3/2})$ of its critical value $\gamma_c = 1/M$ for this evolution to occur asymptotically.

	This $\gamma_c$ and runtime are in agreement with Table~\ref{table:simplex_larger}. 

\subsection{Two Marked Per Complete Graph, Case 2}

	As shown in Figure~\ref{fig:simplex_2M1}b, the system evolves in a 3-dimensional subspace spanned by
	\begin{align*}
		\ket{a} &= \frac{1}{\sqrt{2(M+1)}} \sum_{i \in \text{red}} \ket{i} \\
		\ket{b} &= \frac{1}{\sqrt{(M-4)(M+1)}} \sum_{i \in \text{blue}} \ket{i} \\
		\ket{c} &= \frac{1}{\sqrt{2(M+1)}} \sum_{i \in \text{yellow}} \ket{i}
	\end{align*}
	In this subspace, the search Hamiltonian \eqref{eq:H} is
	\[ H = -\gamma \begin{pmatrix}
		1 + \frac{1}{\gamma} & \sqrt{2(M-4)} & 3 \\
		\sqrt{2(M-4)} & M-4 & \sqrt{2(M-4)} \\
		3 & \sqrt{2(M-4)} & 1 \\
	\end{pmatrix}. \]

	\begin{figure}
	\begin{center}
		\subfloat[] {
			\includegraphics[scale=0.7]{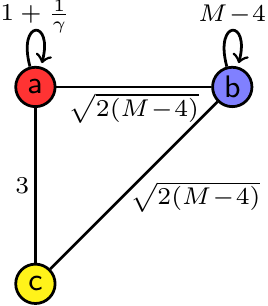}
		} \quad \quad \quad \quad \quad \quad 
		\subfloat[] {
			\includegraphics[scale=0.7]{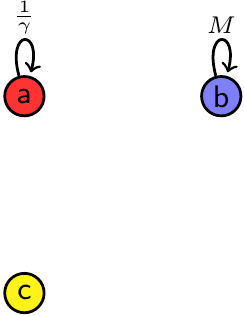}
		}
		\caption{\label{fig:simplex_2M1b_diagrams}Apart from a factor of $-\gamma$, \textbf{(a)} the Hamiltonian for the second case of search on the simplex of complete graphs with $k = 2(M+1)$ marked vertices, and \textbf{(b)} the leading-order terms. }
	\end{center}
	\end{figure}

	This Hamiltonian can be visualized as shown in Figure~\ref{fig:simplex_2M1b_diagrams}a. The leading-order Hamiltonian $H^{(0)}$ excludes edges that scale less than $M$, and it can be visualized as shown in Figure~\ref{fig:simplex_2M1b_diagrams}b. Clearly, the eigenstates of this are $\ket{a}$, $\ket{b}$, and $\ket{c}$ with corresponding eigenvalues $-1$, $-\gamma M$, and $0$. Since $\ket{s} \approx \ket{b}$, we choose $\gamma$ so that $\ket{b}$ and $\ket{a}$ are degenerate, \textit{i.e.},
	\[ \gamma_c = \frac{1}{M} \]
	The perturbation $H^{(1)}$, which restores edges of weight $\Theta(\sqrt{M})$, causes certain linear combinations $\alpha_a \ket{a} + \alpha_b \ket{b}$ to be eigenstates of $H^{(0)} + H^{(1)}$. The coefficients can be found in the usual way, and they yield perturbed eigenstates
	\[ \ket{\psi_0} = \frac{1}{\sqrt{2}} ( \ket{b} + \ket{a} ), \quad E_0 = -1 - \sqrt{\frac{2}{M}} \]
	\[ \ket{\psi_1} = \frac{1}{\sqrt{2}} ( \ket{b} - \ket{a} ), \quad E_1 = -1 + \sqrt{\frac{2}{M}}. \]
	So the system evolves from $\ket{s} \approx \ket{b}$ to $\ket{a}$ in time $\pi / \mathrm{\Delta} E$:
	\[ t_* = \frac{\pi}{2} \sqrt{\frac{M}{2}}. \]

	We can again use the method of \cite{Wong2015e} to find how precisely $\gamma$ must be chosen to its critical value $\gamma_{c} = 1/M$---a straightforward calculation shows that it must be within $o(1/M^{3/2})$.

	This $\gamma_c$ and runtime are in agreement with Table~\ref{table:simplex_larger}. 

\end{document}